\documentclass[12pt]{article}

\usepackage{graphicx}
\usepackage{amsmath}
\usepackage{amsmath,amsfonts}
\usepackage[left=1in, right=1in]{geometry}
\usepackage{epstopdf}
\usepackage{tensor}
\usepackage{slashed}
\usepackage{cite}

\def \be {\begin{equation}}
\def \ee {\end{equation}}
\def \bea {\begin{eqnarray}}
\def \eea {\end{eqnarray}}
\def \nn {\nonumber}

\allowdisplaybreaks

\newcommand{\cJ}{\mathcal{J}}

\newcommand{\sig}{\sigma^{(i)}_{\mu\nu}}
\newcommand{\jProd}{\left ( \prod_{j=1}^{n} J^{f}(k_j) \right) \, }

\newcommand{\KK}{\tensor{K}{^\rho^\sigma_\mu_\nu}}
\newcommand{\GG}{\tensor{G}{^\rho^\sigma_\mu_\nu}}
\newcommand{\TGG}{\tensor{\tilde{G}}{^\rho^\sigma_\mu_\nu}}
\newcommand{\TKK}{\tensor{\tilde{K}}{^\rho^\sigma_\mu_\nu}}

\newcommand{\bpsi}{\overline{\psi}}

\newcommand{\softs}{\underline{s}}

\newcommand{\hl}{\hat{l}}

\newcommand{\qslash}{\slashed{q}}
\newcommand{\pslash}{\slashed{p}}
\newcommand{\kslash}{\slashed{k}}

\begin{document}

\title{Soft Graviton Emission at High and Low Energies in Yukawa and Scalar Theories}

\author{Hualong Gervais \\ C.N.\ Yang Institute for Theoretical Physics\\ and Department of Physics and Astronomy\\
Stony Brook University, Stony Brook, NY 11794-3840 USA}

\begin{flushright}
YITP-SB-17-22
 \end{flushright}

{\let\newpage\relax\maketitle}

\begin{abstract}
We study corrections to the soft graviton theorem at all loop orders in Yukawa and scalar theories, both in the high energy and low energy regions. It is found that the tree level soft theorem is corrected by matrix elements coupled to the Riemann curvature tensor of linearized gravity. Further corrections appear in the high energy region and we develop a power counting technique to classify all such loop corrections according to their order of magnitude.  This leads to the construction of factorized contributions to the soft theorem, to which we apply an analysis analogous to Low's theorem based on the gravitational Ward identity. In this analysis, we emphasize the role played by the external kinematics.
\end{abstract}

\tableofcontents


\section{Introduction} \label{sec:introduction}

The emission of soft particles in a quantum field theory is strongly dependent on the theory's underlying symmetries and classical limit.  In particular, in the limit of zero-energy gauge vector and graviton emission, the leading (1/energy) behavior is determined by dressing the corresponding nonradiative amplitude algebraically with factors representing emissions from external lines.   This result is quite direct at tree level, assuming that the energy of the soft quantum is small compared to all other energy scales.   In certain cases, however, it can extend to all loops.     Low's classic theorem \cite{Low:1958sn} shows that for photon emission, the next level in power corrections is determined by the gauge invariance of the theory, implemented through Ward identites.   The first non-leading contribution in the energy of a soft photon is found from single derivatives of the non-radiative amplitude, so long as these derivatives are well-defined.    The observation in Ref.\ \cite{Cachazo:2014fwa} that at tree level in perturbative quantum gravity, where the leading power for graviton emission has been known for a very long time \cite{Weinberg:1965nx}, the first and second powers in soft graviton emission are found from first and second derivatives of the non-radiative amplitude has stimulated interest in soft theorems in both gravity and gauge theories.  While the Ward identities familiar from gauge invariance are adequate to derive many of these results \cite{Bern:2014vva}, and they may reflect other symmetries as well \cite{Cachazo:2014fwa}, the role of loop corrections, especially in the high-energy and massless limits has received a lot of further study \cite{Gervais:2017yxv,DelDuca:1990gz,  Bern:2014oka,He:2014bga,Cachazo:2014dia, Larkoski:2014bxa}. Other approaches to the study of soft theorems include scattering equations \cite{Cachazo:2013hca,Cachazo:2013iea,Schwab:2014xua}, string theory methods \cite{Geyer:2014lca, Schwab:2014fia, Bianchi:2014gla, Sen:2017xjn}, the one particle irreducible effective action \cite{Sen:2017nim, Laddha:2017ygw}, and path integral and diagrammatic analyses \cite{White:2009aw,White:2011vh, White:2011yy,White:2013ye,White:2014qia,  Laenen:2008gt,Laenen:2010uz,Luna:2016idw}.

In this paper, we focus on loop corrections to the soft graviton theorem, specifically for the emission of a single soft graviton from scalar and Yukawa theories in four dimensions.   We are interested in the high energy limit, studied by Del Duca \cite{DelDuca:1990gz} in the context of soft photon emission in the wide-angle scattering of charged particles, and in low energies.   Ref.\ \cite{DelDuca:1990gz} pointed out that the original form of Low's theorem holds only for photon energies below the scale $m^2/E$, with $m$ the mass scale of virtual lines and $E$ the typical center-of-mass energy of the non-radiative amplitude.   Del Duca showed that for $E_\gamma>m^2/E$, corrections to Low's theorem appear in the first power correction, and that they can be interpreted in terms of infrared-sensitive matrix elements involving the field strength, associated with the collinear singularities in the massless limit.  These contributions are not determined directly by the Ward identities.    They remain universal, however, depending only on the charge, spin, and momentum of the external lines.   This universality is a generalization and variant of the factorization theorems that play such a large role in applications of gauge invariance in perturbative quantum chromodynamics \cite{Collins:1989gx,Bonocore:2015esa}. In the case of gravity, we will find analogous corrections where the role of Del Duca's field strength is played by the Riemann tensor of linearized gravity, both at high and low energies -- see Eq.\ \eqref{eq:Riemanncoupling} below.\footnote{After this work was completed, Ref.\ \cite{Laddha:2017ygw} appeared, which also derives corrections that we identify as the linearized Riemann tensor.}

The full soft graviton theorem \cite{Cachazo:2014fwa,Weinberg:1965nx,Bern:2014vva} applies to an $n+1$-point  amplitude $\mathcal{M}_{n+1}(k_1,...,k_n,q)$ where the $k_1,...,k_n$ are hard momenta with $k_i \cdot k_j >> k_i^2, k_j^2$ for all $i,j$, and $q$ is a soft graviton momentum. In the limit that $q^\mu$ vanishes in all components relative to all $k_i \cdot k_j$,  we can expand in $q$, starting with the leading, $1/q$, behavior
\begin{align}
\mathcal{M}_{n+1}(k_1,...,k_n,q) &= (S_0+S_1+S_2)\mathcal{M}_n(k_1,...,k_n)\ +\ O( q^2) \, ,
\label{eq:CSresult}
\end{align}
where the $S_i$ specify in closed form the leading and the first two subleading power corrections in the graviton momentum, $q^\mu$,
\begin{align}
S_0 &= \sum_{i=1}^{n}\frac{E_{\mu\nu}k_i^\mu k_i^\nu}{k_i\cdot q} \nonumber\\
S_1 &= \sum_{i=1}^{n}\frac{E_{\mu\nu}k_i^\mu(q_\rho \cJ_{i}^{\rho\nu})}{k_i\cdot q} \nonumber\\
S_2 &= \frac{1}{2}\sum_{i=1}^{n}\frac{E_{\mu\nu}(q_\rho \cJ_{i}^{\rho\mu})(q_\sigma \cJ_{i}^{\sigma\nu})}{k_i\cdot q}\, .
\label{eq:S012}
\end{align}
Here $E_{\mu\nu}$ is the soft graviton polarization tensor and $\mathcal{J}_i^{\mu\nu}$ is the angular momentum tensor of the $i^{th}$ external particle, of the form 
\begin{align} 
\cJ_i^{\mu\nu} \equiv k_i^\mu \frac{\partial}{\partial k_{i\nu}} - k_i^\nu \frac{\partial}{\partial k_{i\mu}}\ +\ \Sigma_i^{\mu\nu}\, ,
\label{eq:orbital}
\end{align}
with $\Sigma^{\mu\nu}$ a spin term. Newton's constant, $\kappa$, has been normalized so that $\kappa/2 = 1$, and we will also make this choice whenever convenient.

In \cite{Cachazo:2014fwa}, Cachazo and Strominger proved that Eqs.\ (\ref{eq:CSresult})-(\ref{eq:orbital}) apply for arbitrary $n$ at tree level using the BCFW construction \cite{Britto:2004ap, Britto:2005fq}.  We will refer to their result as ``CS'' below. Subsequently, the CS result was rederived from the gravitational Ward identity \cite{Weinberg:1964ew} that decouples scalar-polarized gravitational radiation \cite{Bern:2014vva}, following the analysis of Low \cite{Low:1958sn} for soft photon radiation in Quantum Electrodynamics.   Refs.\ \cite{Bern:2014vva,Bern:2014oka} discussed modifications associated with loop corrections and soft singularities, both for pure gravity and for gravitational radiation associated with massive and massless matter fields.

Our approach to loop corrections at high energies closely follows Ref.\ \cite{Gervais:2017yxv}, which revisits the problem of photon emission at high energies studied by del Duca. Ref.\ \cite{Gervais:2017yxv} focuses on Yukawa and scalar theories in the regime where the soft momentum $q$ is of the order $O(m^2 /E)$. 
In Ref\  \cite{Gervais:2017yxv}, it was pointed out that in the result of loop integration over regions neighboring pinch surfaces, there arise contributions with branch cuts within $O(m^2/E)$ of the point $q = 0$. Following Ref.\ \cite{Gervais:2017yxv}, we will refer to such contributions as ``non-analytic''. Branch cuts of non analytic contributions are associated with particle production thresholds and make an expansion of the non-radiative amplitude (henceforth referred to as the ``elastic'' amplitude) to a fixed power of $q$ inaccurate when $q = O(m^2 /E)$. This phenomenon can be traced back to invariants involving $q$ being of the same order as terms not involving $q$ in denominators of the loop integrand \cite{Gervais:2017yxv, DelDuca:1990gz}. Expanding the elastic amplitude, however, is a key step in Low's original analysis -- see \cite{Low:1958sn, Bern:2014vva} for details. 

A solution to this problem is to factorize the elastic amplitude into jet functions, a soft cloud, and a hard part. The jet functions gather all collinear lines of the diagram, the soft cloud gathers all the soft lines, and the hard part includes all hard exchanges. Such a factorization is reminiscent of the soft collinear effective theory approach to soft radiation \cite{Larkoski:2014bxa}. The hard part is analogous to the matching coefficients of SCET and the jet functions correspond to the higher dimension operators of the effective theory. Ref.\ \cite{Gervais:2017yxv} applied power counting techniques to provide a systematic way of classifying factorized contributions to the radiative amplitude according to their order of magnitude. This allowed for a complete list of all non-leading loop corrections to the soft photon theorem, carefully taking into account the analytic structure of the loop integrals in all regions. At high energies, this paper applies the same technique to the soft graviton theorem to obtain a complete classification of all higher loop corrections to the CS result. Further, we will also provide a decomposition of the radiative jet functions inspired from Grammer and Yennie's decomposition \cite{Grammer:1973db} and find that loop corrections sensitive to the collinear region couple to the graviton through a linearized Riemann tensor, both at high and low energies.

This paper is organized as follows. Section \ref{subsec:power_counting} focuses on the elastic amplitude and outlines the main steps of our power counting technique to derive all reduced elastic diagrams contributing to the soft graviton theorem at high energies. Section \ref{subsec:elastic_factorization} lists all factorized amplitudes corresponding to the reduced diagrams previously identified. Section \ref{subsec:graviton_emission_power_counting} introduces the subject of graviton emission proper by first treating how the result of a power counting analysis of the radiative amplitude is related to the analogous power counting for the elastic amplitude. Section \ref{subsec:low_analysis} then discusses how Low's analysis is extended to a factorized amplitude involving leading jets only, with emphasis on a careful transition between radiative and elastic kinematics. The extension of Low's analysis to non-leading jets is described in Sec.\ \ref{subsec:graviton_emission_non_leading}. The low energy limit of the soft theorem is treated in Sec.\ \ref{subsec:graviton_low_energy}.  Section \ref{subsec:kg_decomposition} concerns graviton emission from the jets and shows how it can be decomposed into a leading contribution and a gauge invariant subleading correction that couples to gravitons through a linearized Riemann tensor. Finally, Sec.\ \ref{subsec:off_shell_emission} concludes with an example of non-leading corrections to an amplitude where an off-shell graviton is exchanged between a jet and a very heavy source.


\section{Expansion of the Elastic Amplitude} \label{sec:elastic_amplitude_expansion}

Pinch surfaces of a loop integral are surfaces in momentum space where the denominator of the integrand vanishes, and that cannot be avoided by rerouting the loop integration contours in the complex plane without crossing a pole. Consequently, in general, non-analytic contributions will arise from loop integration regions close to pinch surfaces. Not all such contributions are singular, however, since the vanishing of denominators can be damped by the vanishing of the numerator. Power counting techniques have been developped to quantify when and how this damping can occur in Minkowski space \cite{Sterman:1978bi,Sterman:1978bj,Libby:1978qf}. 

Although non-analytic terms arising from pinch surfaces may not always produce singularities, they may not be expandable as a power series in the soft momentum $q$. As explained in \cite{Gervais:2017yxv}, this is important in the context of the soft theorem since the core of Low's argument relies on using the Ward identity to introduce a $q$ dependence into the elastic amplitude to derive the emission amplitude from internal lines. The $q$ dependent elastic amplitude is then expanded in $q$ using a Taylor series. At high energies, with $q\sim m^2/E$, the presence of pinch surfaces is an obstacle to the application of Low's analysis since their associated non-analytic terms may not be accurately expanded about $q = 0$. For an expansion about $q= 0$ to be valid, we would need the soft momentum to be demoted to a region where $q << m^2/E$. It is therefore necessary to factorize these non-analytic terms so that the radiative amplitude is factored into terms that cannot be expanded in $q$ (the jets functions and soft cloud) and terms that can (the hard parts).

To construct a factorized graviton emission amplitude, we will first derive a factorized elastic amplitude and then consider all lines and vertices from which the soft graviton may be emitted. Let $\mathcal{M}_{el}(k_1, \dots, k_n)$ be an elastic scattering amplitude, where for convenience we take all external particles to be outgoing. In this section, the momenta $k_1, \dots, k_n$ are assumed to belong to fermions or antifermions for simplicity, but the extension of our arguments to external scalars is straightforward. We assume that the fermions have a non vanishing mass $m$ while the scalars are kept massless. Further, let $E$ be the ``large'' energy scale set by the invariants built from the external momenta. The fermion mass is used to quantify the meaning of ``small''. Concretely, we define the parameter $\lambda \equiv m/E$ and use it to express the order of magnitude of any quantity. Given an arbitrary quantity $F$, the equation $F = O(\lambda^\gamma)$ means that there exists a constant $f$ such that $|F| < f \lambda^\gamma$ where $f$ is a positive real number times an appropriate power of $E$ to match the dimension of $F$. It is crucial that $f$ does not depend on $m$  in this definition.

In the high energy regime, $\lambda$ is small and we have that the soft graviton momentum scales as $q = O(\lambda^2)$. The soft graviton theorem can then be thought of as an expansion of the graviton radiative amplitude in powers of $\lambda$,
\begin{align}\label{eq:soft_theorem_lambda_expansion}
\mathcal{M}^{\mu\nu}(k_1, \dots, k_n, q) &= \sigma^{\mu\nu}_{-2} + \sigma^{\mu\nu}_{-1} + \sigma^{\mu\nu}_{0} + \sigma^{\mu\nu}_{1} + \sigma^{\mu\nu}_{2} \, ,
\end{align}
where $\sigma^{\mu\nu}_{\gamma} = O(\lambda^\gamma)$. Note that in the CS result, only even $\gamma$ terms are present. As we will demonstrate, this is no longer the case when we consider loop corrections at high energies.

Our first task will be to find a similar expansion in powers of $\lambda$ of the elastic amplitude. For reasons that will become clear in Sec.\ \ref{subsec:graviton_emission_power_counting}, it is necessary for us to consider the expansion of $\mathcal{M}_{el}(k_1, \dots, k_n)$ up to $O(\lambda^4)$. To derive such an expansion of a loop integral contributing to $\mathcal{M}_{el}(k_1, \dots, k_n)$, we first locate all pinch surfaces of the integrand by solving the Landau equations \cite{Landau:1959fi}. Solutions to these equations can be visualized using reduced diagrams that represent the classical propagation of on-shell particles \cite{Coleman:1965xm}. As shown in detail in \cite{Gervais:2017yxv}, one can use power counting techniques \cite{Sterman:1978bi, Sterman:1978bj,Sterman:1994ce, Sterman:1995fz, Collins:2011zzd} to determine the order in $\lambda$ of the loop integral when the integration domain is restricted to a region neighboring a pinch surface. Therefore, one can separate the whole range of a loop integral into regions surrounding the pinch surfaces of the integrand, with each pinch surface yielding a factorized contribution of order $\lambda^\gamma$, where $\gamma$ is the infrared degree of divergence of the loop integral near that  pinch surface.

\subsection{Power Counting} \label{subsec:power_counting}

We proceed in deriving an expansion of the elastic amplitude by determining which pinch surfaces have degree of divergence $\gamma \leq 4$. As in \cite{Gervais:2017yxv}, this will be done by applying power counting techniques to the reduced diagrams representing the pinches. Our methodology is closely related to the power counting procedures presented in \cite{Akhoury:1978vq, Sen:1982bt, Sterman:1995fz}.  The power counting rules relevant to Yukawa and scalar theory with massive fermions and massless scalars are shown in Table \ref{tab:power_counting}.

To make our conventions clear, we repeat a few definitions from \cite{Gervais:2017yxv} that we will also use. Given the external momentum $k_i$, we can define collinear and anti-collinear vectors $n_i$ and $\bar{n}_i$ by the requirements
\begin{align}
 \vec{n}_i &= \frac{\vec{k}_i}{\sqrt{2} |\vec{k}_i|} = - \vec{\bar{n}}_i \nonumber \, , \\ 
 n_i \cdot \bar{n}_i &= 1 \, , \nonumber \\
 n_i^2 &= \bar{n}_i^2 = 0 \, .
\end{align}
The light cone coordinates of an arbitrary vector $v$ relative to the external momentum $k_i$ are then defined by 
\begin{align}
v &=  v^+ n_i +  v^- \bar{n}_i + v_T \, ,
\end{align}
where $v_T$ is the transverse vector defined such that $v_T^0 = 0$ and $v_T \cdot k_i = 0$. It is customary to denote a vector $v$ by its light cone coordinates $(v^+, v^-, v_T)$.

A momentum $l$ collinear to $k_i$ is defined to scale as $l \sim (1, \lambda^2, \lambda) E $. On the other hand, a soft momentum scales as $l \sim (\lambda^2, \lambda^2, \lambda^2) E$ and finally a hard momentum scales as $l \sim (1, 1, 1) E$.


\begin{table}[h!]
\caption{The power counting rules below define how much each component of a reduced diagram contributes to the degree of divergence of the corresponding pinch surface. These rules are for Yukawa and scalar theories where the fermions are massive and the scalars are massless. In the case of massless fermions, soft fermions yield an enhancement of $-2$ rather than $-1$. }

\begin{center}
	\begin{tabular}{ccc} 
	\hline
	\hline
	                                                                 &    Enhancement    &     Suppression \\ \hline
	Collinear fermion line                              &        -2                   &                          \\
	Collinear scalar line                                &        -2                  &                           \\
	Soft fermion line                                     &        -1                &                            \\
	Soft scalar line                                        &         -4                 &                           \\
	Collinear loop integral                              &                              &           +4           \\
	Soft loop integral                                     &                              &           +8             \\
	Yukawa vertex on collinear fermion line  &                              &            +1            \\\hline \hline
	\end{tabular}

\label{tab:power_counting}
\end{center}
\end{table}

\begin{figure}[h!]
\centering
\includegraphics[width = 0.75 \textwidth]{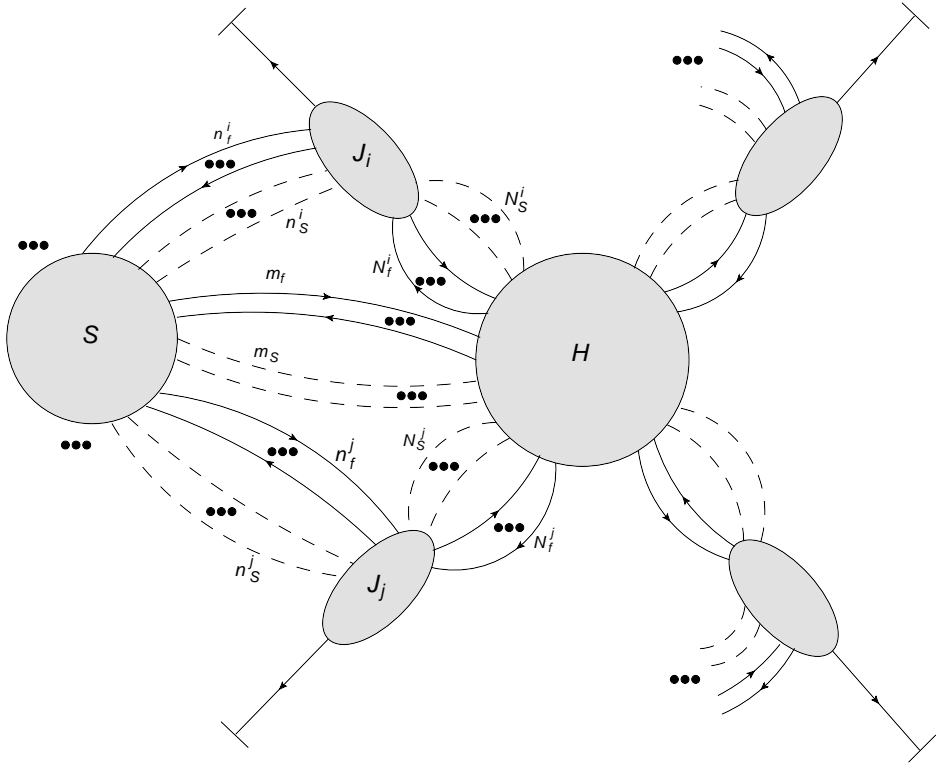}
\caption{The most general reduced diagram incorporating the hard vertex, the soft function, and well separated jets.}
\label{fig:generic_multi_jet}
\end{figure}

Using the Euler identity, one can show that the degree of divergence $\gamma$ of the most general reduced diagram (shown in Fig.\ \ref{fig:generic_multi_jet}), is given by the formula \cite{Gervais:2017yxv}
\begin{align} \label{eq:power_counting_final_result}
\gamma &= \sum_i (N^i_f + N^i_s + 3 n^i_f + n^i_s -1) + I_f + 4 m_f + 2 m_S \, .
\end{align}
In the above, $N^i_f$ and $N^i_s$ stand respectively for the number of collinear fermions and scalars emerging from the hard part into the $i^{th}$ jet. Likewise, $n^i_f$ and $n^i_s$ denote the number of soft fermions and scalars connecting the $i^{th}$ jet to the soft cloud. Finally, $I_f$ is the number of soft fermions internal to the soft cloud, while $m_f$ and $m_S$ are the number of soft fermions and scalars attaching the soft cloud to the hard part.


It is convenient to define the quantities
\begin{align} \label{eq:gamma_definitions}
\tilde{\gamma} & \equiv \gamma  - I_f - 4 m_f - 2 m_S  \nonumber \\
 &=\sum_i ( N^i_f + N^i_s + 3 n^i_f + n^i_s -1) \equiv \sum_i \gamma_i \, .
\end{align}
One can think of $\gamma_i$ as an effective contribution to the degree of divergence from the $i^{th}$ jet after the effect of the soft cloud has been taken into account. Finding all diagrams with $ 0 \leq \gamma \leq 4$ can be accomplished by first searching for all reduced diagrams with $0 \leq \tilde{\gamma} \leq 4$ and then enforcing the necessary constraints on $I_f$, $m_f$, and $m_S$.

\begin{figure}[h!]
\centering
\includegraphics[width = 0.75 \textwidth]{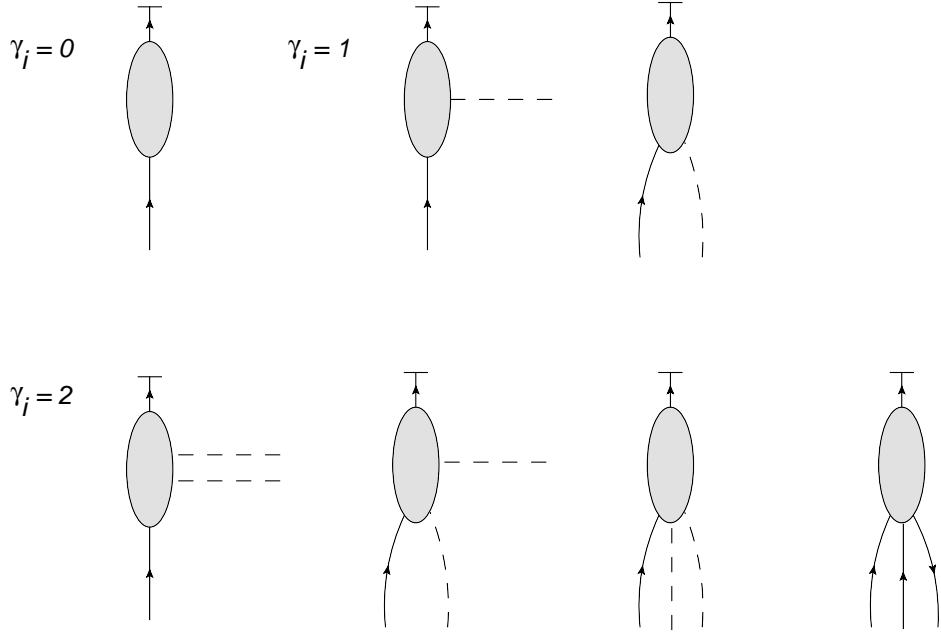}
\caption{These jets have effective degree of divergence $\gamma_i \leq 2$.}
\label{fig:gamma_i_classification_1}
\end{figure}

\begin{figure}[h!]
\centering
\includegraphics[width = 0.75 \textwidth]{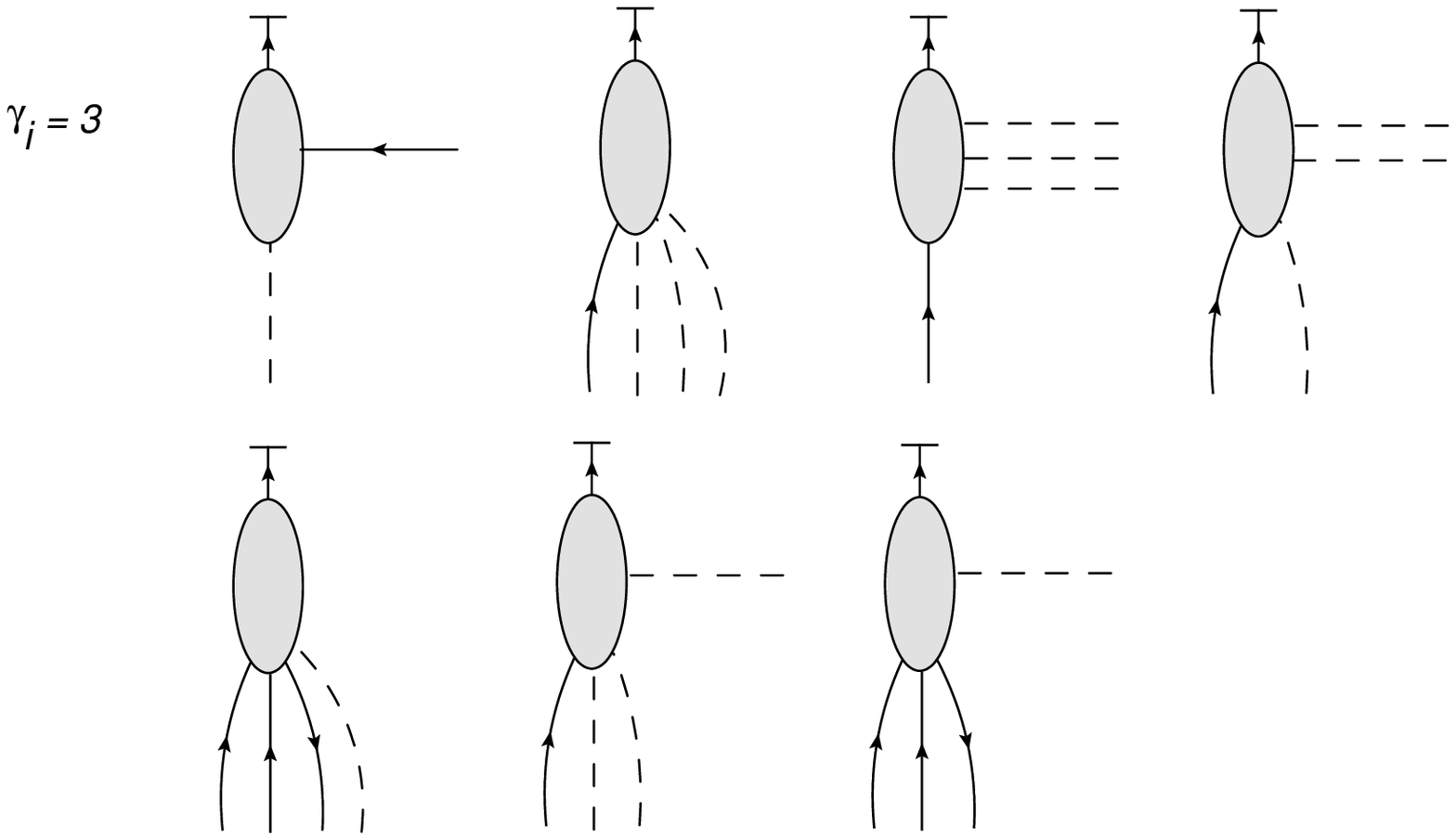}
\caption{These jets have effective degree of divergence $\gamma_i = 3$.}
\label{fig:gamma_i_classification_2}
\end{figure}

\begin{figure}[h!]
\centering
\includegraphics[width = 0.75 \textwidth]{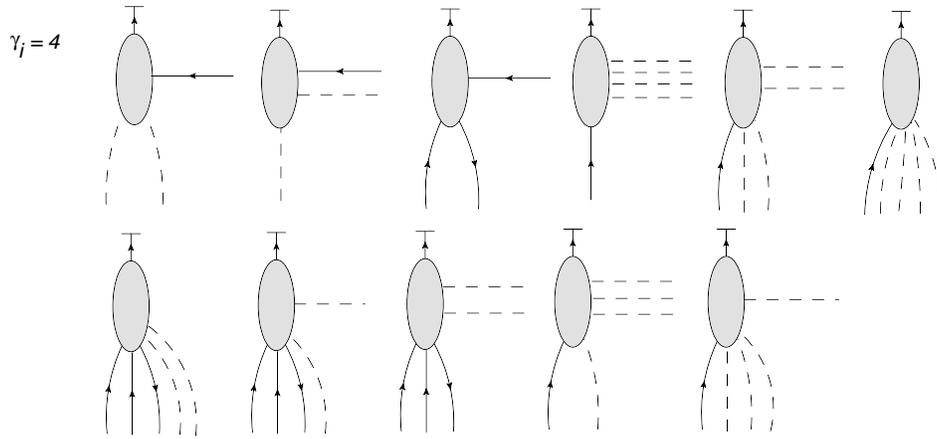}
\caption{These jets have effective degree of divergence $\gamma_i = 4$.}
\label{fig:gamma_i_classification_3}
\end{figure}

The first step in identifying all reduced diagrams with $0 \leq \tilde{\gamma} \leq 4$ is to determine all jets with $\gamma_i = 0,1,2,3,$ or $4$. The results follow from an inspection of the $\gamma_i$, defined in \eqref{eq:gamma_definitions} and shown in Figs.\ \ref{fig:gamma_i_classification_1}, \ref{fig:gamma_i_classification_2}, and \ref{fig:gamma_i_classification_3}.


From the definition $\tilde{\gamma} = \sum_i \gamma_i$, it is then clear that to find all reduced diagrams with a given value of $\tilde{\gamma}$, we need to find first all distinct partitions of that value into a sum of positive integers. Then, for each of the summands, we need to choose one jet with matching $\gamma_i$. For example, suppose we want to find a reduced diagram with $\tilde{\gamma} = 4$. One of the partitions of $4$ is $4= 2+1 +1$. We then need one jet with $\gamma_i = 2$  and two jets with $\gamma_i = 1$ from Fig.\ \ref{fig:gamma_i_classification_1}.

\begin{figure}[h!]
\centering
\includegraphics[width = 0.75 \textwidth]{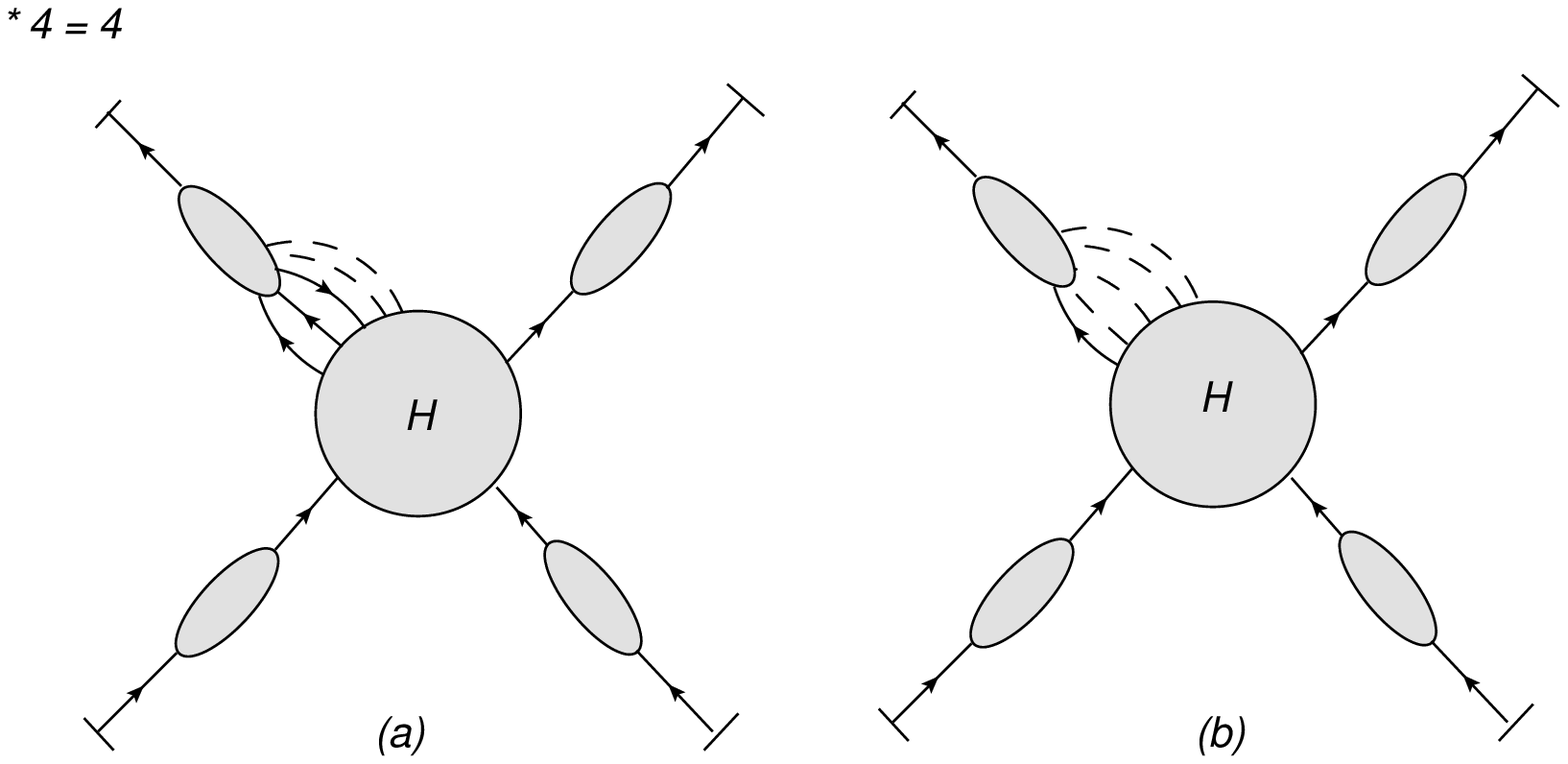}
\caption{These diagrams have $\tilde{\gamma} = 4$ and correspond to the trivial partition $4 = 4$.}
\label{fig:g_4_partition_4}
\end{figure}

\begin{figure}[h!]
\centering
\includegraphics[width = 0.75 \textwidth]{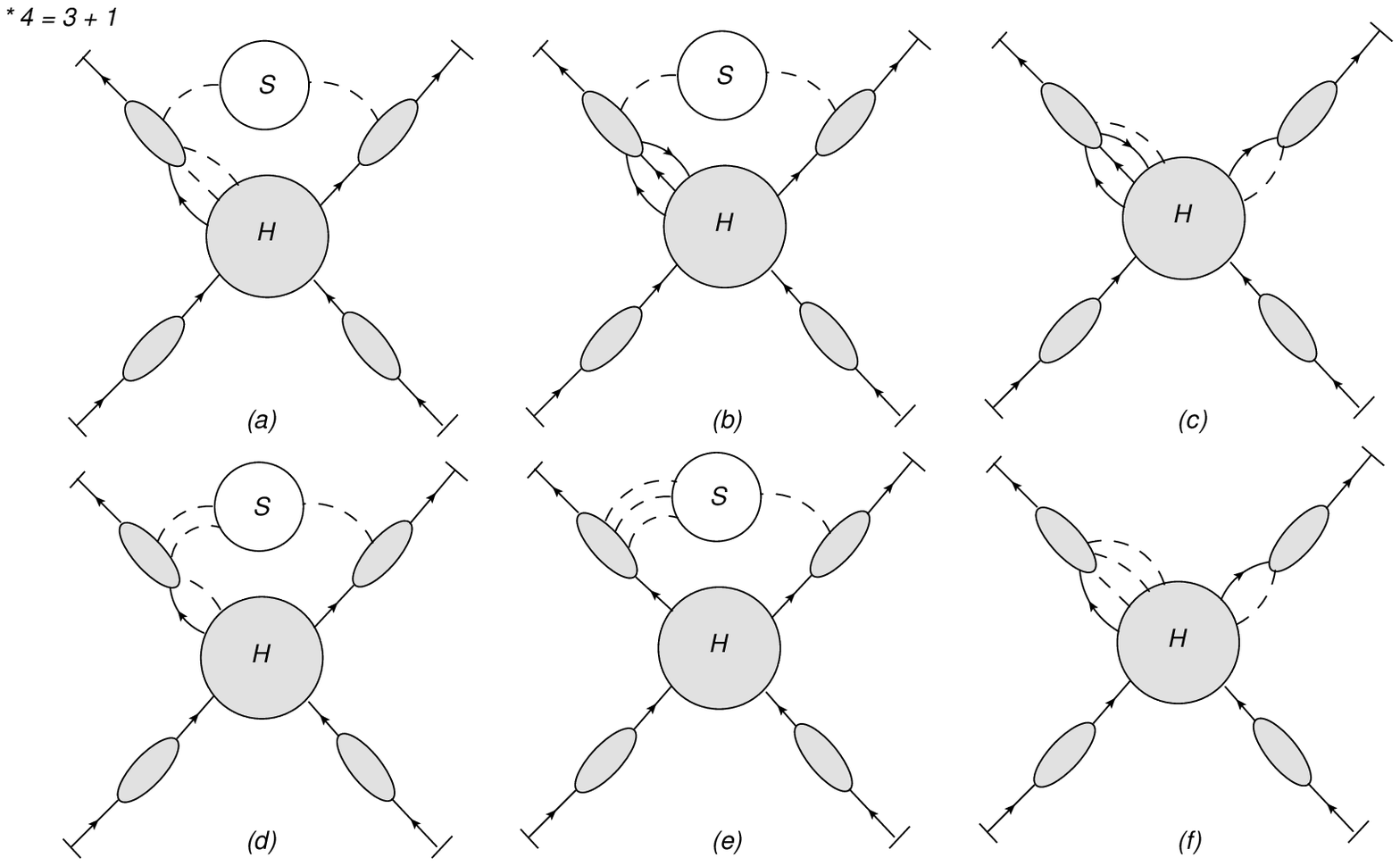}
\caption{These diagrams have $\tilde{\gamma} = 4$ and correspond to the partition $4 = 3 + 1$. Diagram (d) vanishes in $\phi^4$ theory since the soft cloud has three scalars emerging from it.}
\label{fig:g_4_partition_31}
\end{figure}

\begin{figure}[h!]
\centering
\includegraphics[width = 0.75 \textwidth]{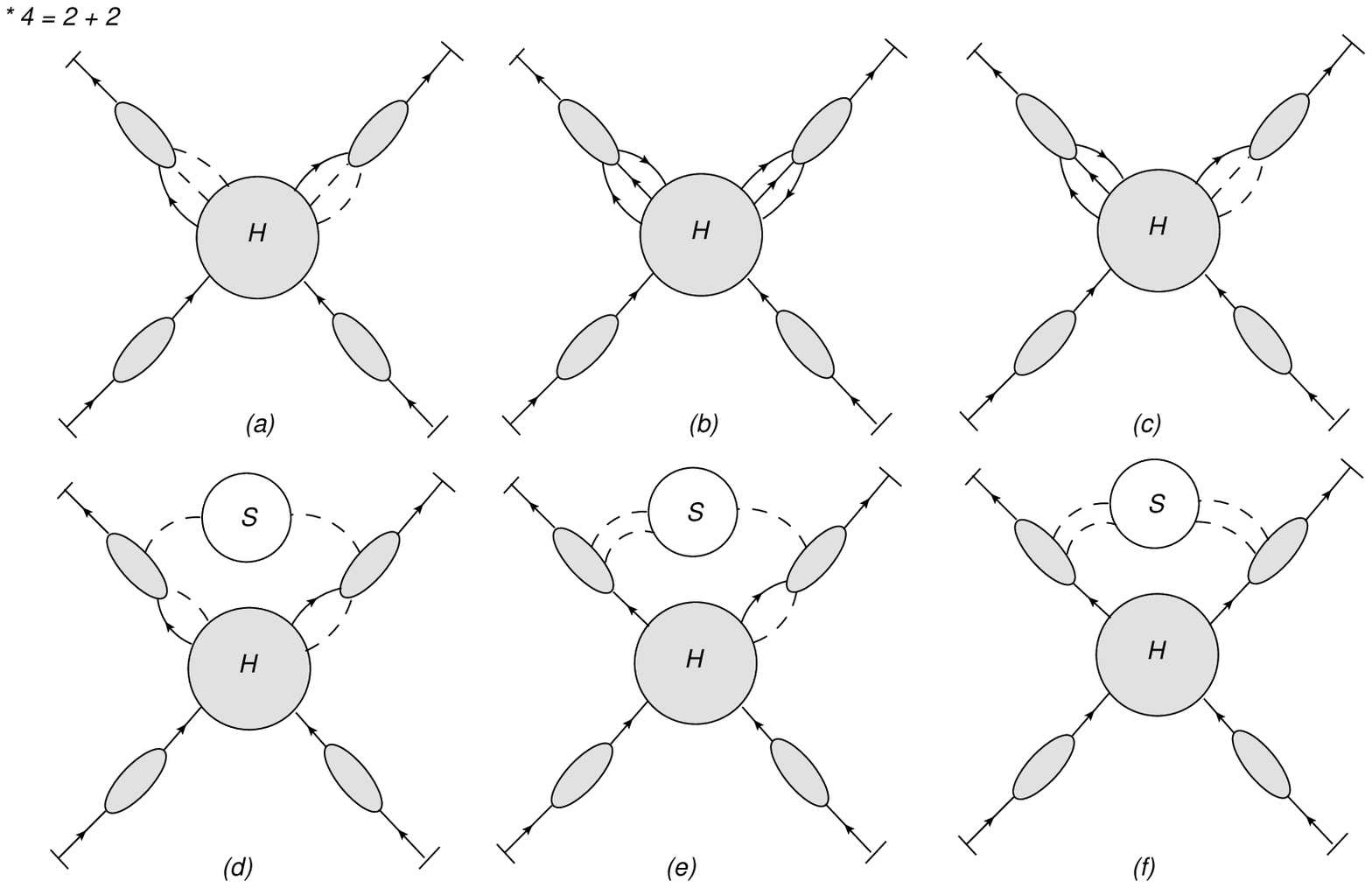}
\caption{These diagrams have $\tilde{\gamma} = 4$ and correspond to the partition $4 = 2 + 2$. Diagram (e) vanishes in $\phi^4$ theory since the soft cloud has three scalars emerging from it.}
\label{fig:g_4_partition_22}
\end{figure}

\begin{figure}[h!]
\centering
\includegraphics[width = 0.75 \textwidth]{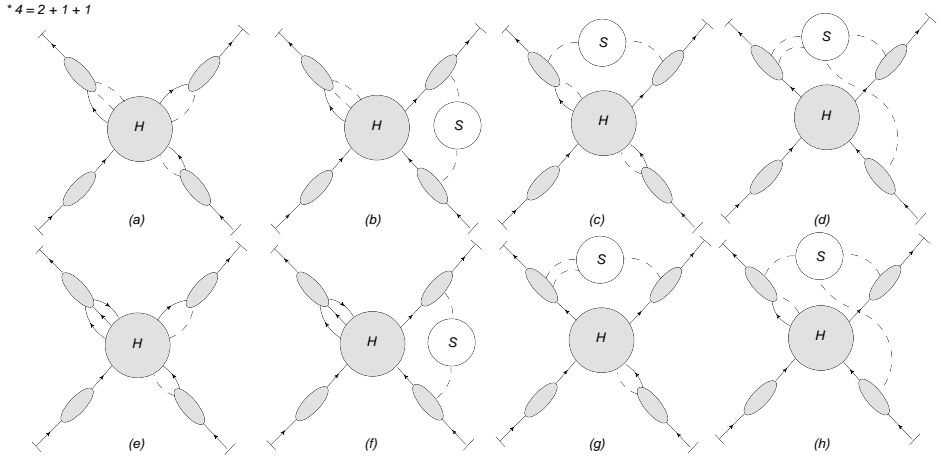}
\caption{These diagrams have $\tilde{\gamma} = 4$ and correspond to the partition $4 = 2 + 1 + 1$. Diagrams (g) and (h) vanish in $\phi^4$ theory since their soft clouds have three scalars emerging from them.}
\label{fig:g_4_partition_211}
\end{figure}

\begin{figure}[h!]
\centering
\includegraphics[width = 0.75 \textwidth]{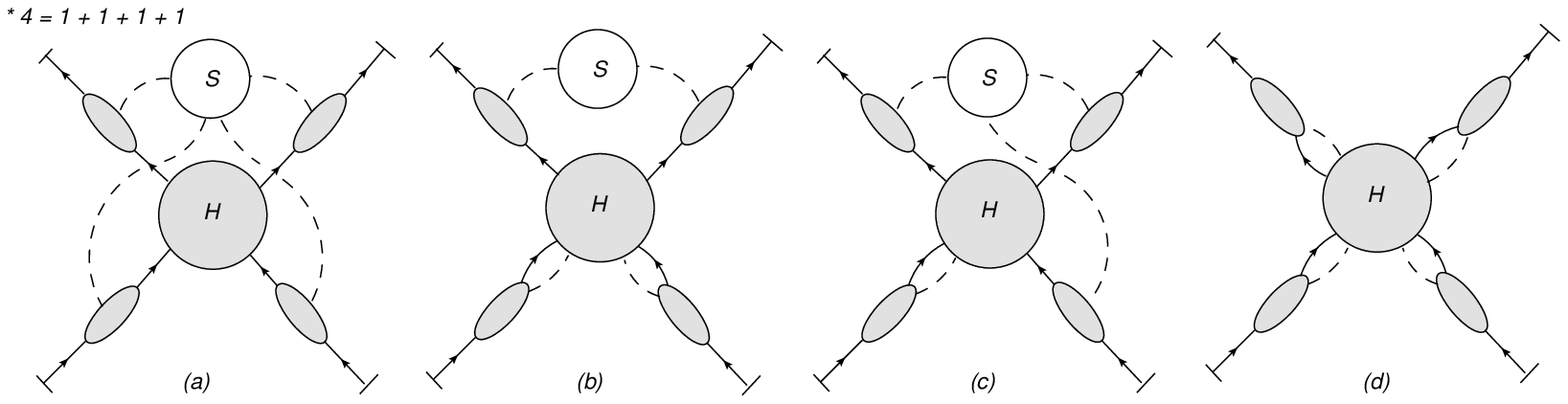}
\caption{These diagrams have $\tilde{\gamma} = 4$ and correspond to the partition $4 = 1 + 1 + 1 + 1$. Diagram (c) vanishes in $\phi^4$ theory since the soft cloud has three scalars emerging from it.}
\label{fig:g_4_partition_1111}
\end{figure}

\begin{figure}[h!]
\centering
\includegraphics[width = 0.75 \textwidth]{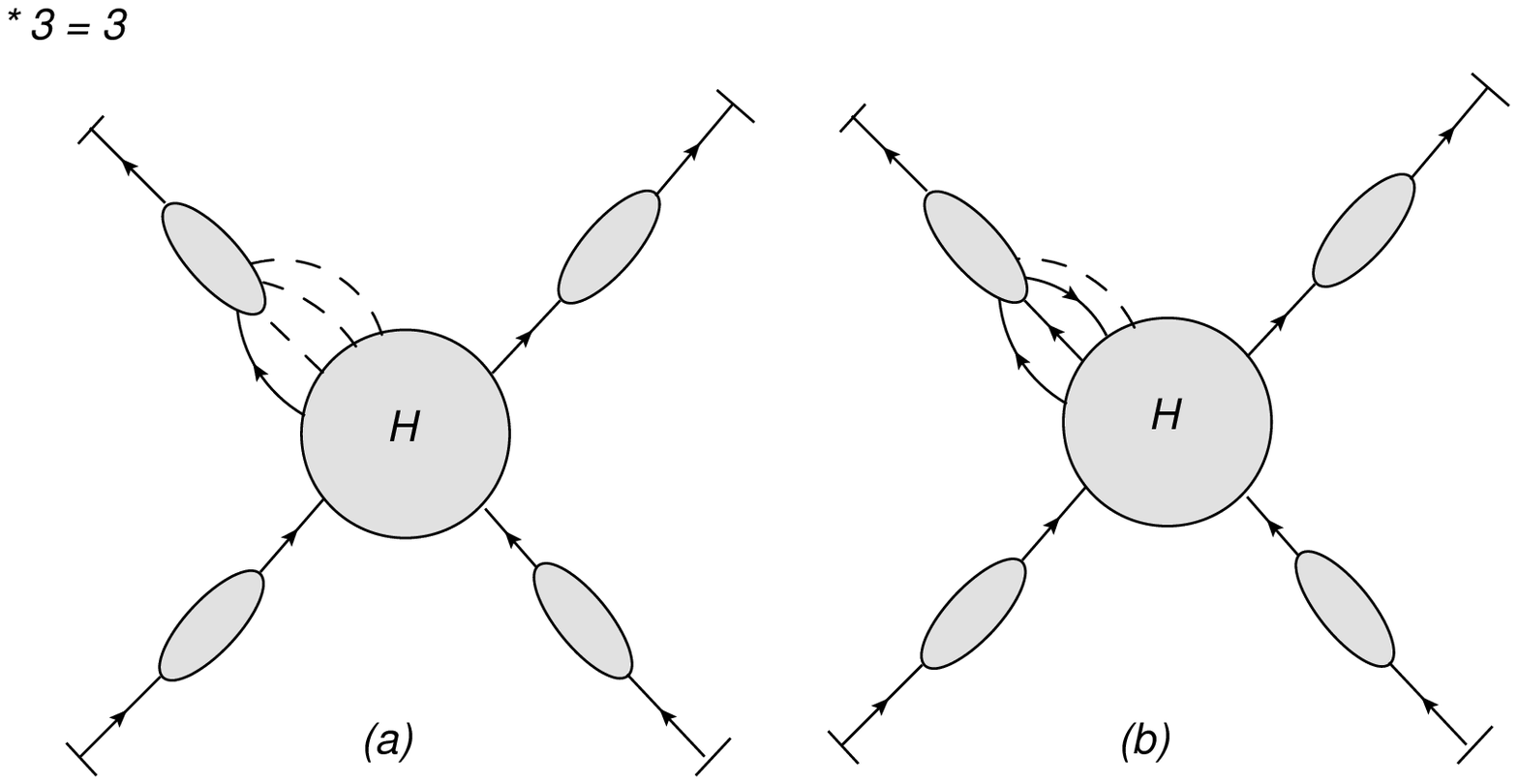}
\caption{These diagrams have $\tilde{\gamma} = 3$ and correspond to the trivial partition $3 = 3$.}
\label{fig:g_3_partition_3}
\end{figure}

\begin{figure}[h!]
\centering
\includegraphics[width = 0.75 \textwidth]{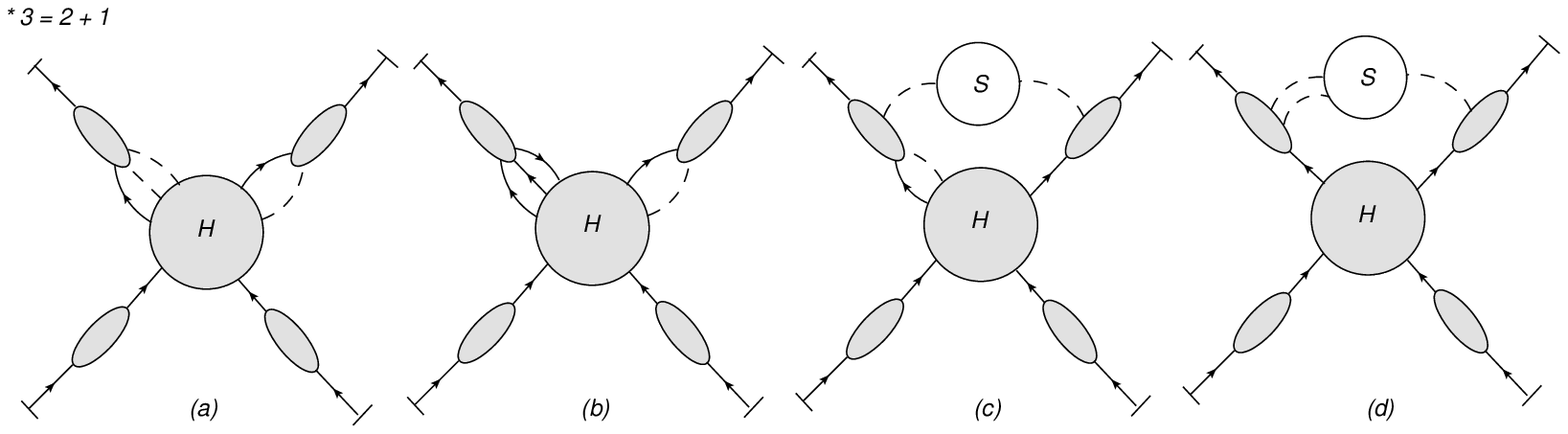}
\caption{These diagrams have $\tilde{\gamma} = 3$ and correspond to the partition $3 = 2 + 1$. Diagram (d) vanishes in $\phi^4$ theory since the soft cloud has three scalars emerging from it.}
\label{fig:g_3_partition_21}
\end{figure}

\begin{figure}[h!]
\centering
\includegraphics[width = 0.75 \textwidth]{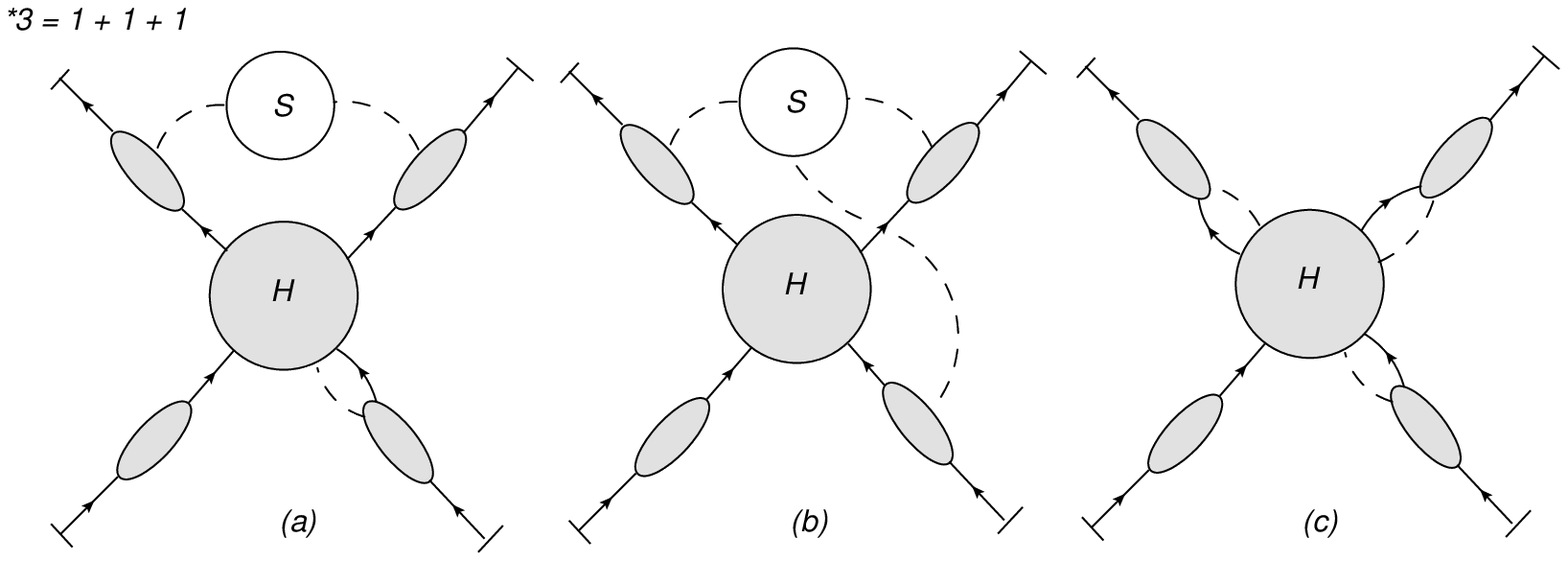}
\caption{These diagrams have $\tilde{\gamma} = 3$ and correspond to the partition $3 = 1 + 1 + 1$. Diagram (b) vanishes in $\phi^4$ theory since the soft cloud has three scalars emerging from it.}
\label{fig:g_3_partition_111}
\end{figure}

To find all reduced diagrams with $\tilde{\gamma} = 4$, we first write down all partitions of $4$,
\begin{align}
\tilde{\gamma} &= 4 \nonumber\\
&= 3+1 \nonumber \\
&= 2+2 \nonumber \\
&= 2+1+1 \nonumber \\
&= 1+1+1+1\, .
\end{align}
The classes of diagrams with $\tilde{\gamma} = 4$ corresponding to each partition are shown in Figs.\ \ref{fig:g_4_partition_4}, \ref{fig:g_4_partition_31}, \ref{fig:g_4_partition_22}, \ref{fig:g_4_partition_211}, and \ref{fig:g_4_partition_1111}. The same approach yields all diagrams with $\tilde{\gamma} = 3$. These are labelled by their corresponding partition of $3$ and shown in Figs.\ \ref{fig:g_3_partition_3}, \ref{fig:g_3_partition_21}, and \ref{fig:g_3_partition_111}. The diagrams with $\tilde{\gamma} \leq 2$ can be found in \cite{Gervais:2017yxv}.


We may now return to the original definition of the degree of divergence $\gamma \equiv \tilde{\gamma} + I_f + 4 m_f + 2 m_s$. In contrast to the soft photon theorem \cite{Gervais:2017yxv}, the need to consider diagrams with degree of divergence up to $\gamma = 4$ also brings about the possibility of having soft lines connecting the soft cloud to the hard part. 

Suppose we consider a diagram with $\tilde{\gamma} = 4$, then the constraint $\gamma \leq 4$ forces us to have $I_f = m_f = m_s = 0$ and $\gamma = \tilde{\gamma} = 4$ in this case. If $\tilde{\gamma} = 3$, then $m_f = m_s = 0$, but we may have $I_f = 0$ of $1$. The former case gives us diagrams with $\gamma = 3$, which are identical to those shown in Figs.\ \ref{fig:g_3_partition_3}, \ref{fig:g_3_partition_21}, and \ref{fig:g_3_partition_111}. The latter case allows us to have exactly one fermion ring with no scalars attaching to it. This disconnected piece can be renormalized to $0$ and therefore we may ignore it.

\begin{figure}[h!]
\centering
\includegraphics[width = 0.8 \textwidth]{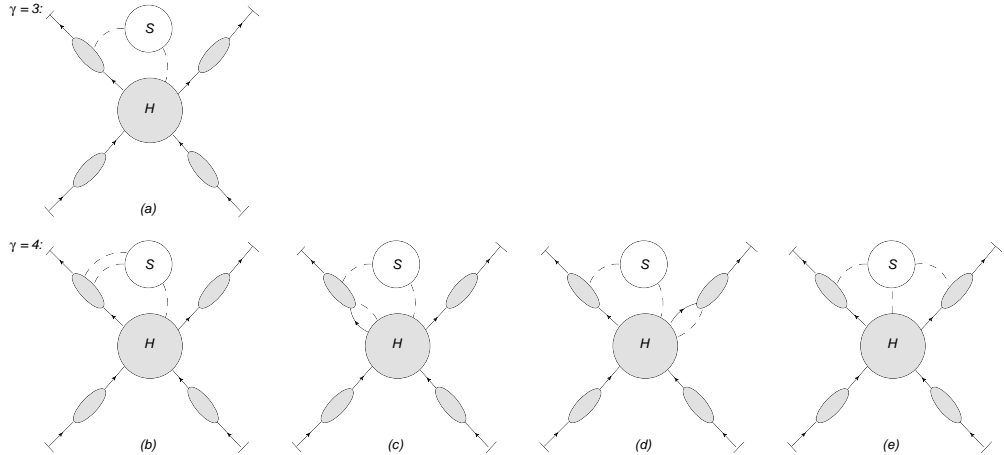}
\caption{The ``exceptional'' diagrams with a soft scalar connecting the soft cloud to the hard part. Diagrams (b) and (e) vanish in $\phi^4$ theory since their soft clouds have three scalars emerging from them.}
\label{fig:hard_connection_example}
\end{figure}

The situations where $\tilde{\gamma} = 0$, $1$, or $2$ give us the freedom to have soft fermions internal to the soft cloud, or soft fermions and scalars connecting the soft cloud to the hard part. For example, if $\tilde{\gamma} = 1$, we may have $\gamma = 3$ and $m_s = 1$. The resulting diagram is shown in Fig.\ \ref{fig:hard_connection_example} (a). The other possibilities are also shown in Fig.\ \ref{fig:hard_connection_example} and we will refer to these diagrams as ``exceptional'' diagrams. Of course,  when $I_f = m_f = m_S = 0$, we have that $\gamma = \tilde{\gamma}$. Consequently, the diagrams shown in Figs.\ \ref{fig:g_4_partition_4} to \ref{fig:g_3_partition_111} all have a degree of divergence that matches their $\tilde{\gamma}$ label. Finally, we remark that diagrams containing a soft cloud with an odd number of external soft scalars vanish in $\phi^4$ theory and, therefore, may be ignored in our analysis.


\subsection{Factorization of the Elastic Amplitude} \label{subsec:elastic_factorization}

Each reduced diagram listed in Figs. \ref{fig:g_4_partition_4} to \ref{fig:hard_connection_example} corresponds to a factorized contribution to the elastic amplitude. The effect of the jets of collinear lines emerging from the hard parts are captured by jet functions defined in terms of matrix elements of the basic field operators of our theory.

A leading jet is essentially a reduced self energy defined by the matrix element \cite{Gervais:2017yxv},
\begin{align}
J^f (k_i) &= \langle k_i | \bpsi(0) | 0 \rangle \, .
\end{align}
The superscript $f$ refers to an outgoing fermion or antifermion and therefore could stand for either $f$ or $\bar{f}$. This jet function corresponds to the situation where a single parton emerges from the hard part, interacts with itself, and propagates to infinity. The case where several collinear particles emerge from the hard part and combine into a single outgoing parton, which may or may not exchange soft quanta with other jets, is referred to as a non-leading jet. We give a few examples of the operator definition of some non-leading jets,
\begin{align}\label{eq:jet_def_examples}
J^{fs}(k_i - \hl, \hl)  &= \int_{-\infty}^{\infty} d\xi \, e^{-i \hl\cdot (\xi \bar{n}_i)} \langle k_i | \phi(\xi \bar{n}_i) \bpsi(0) | 0 \rangle \, , \nonumber \\
J^{fss} (k_i - \hl_1 - \hl_2, \hl_1, \hl_2) &= \int_{-\infty}^{\infty} d\xi_1 \, e^{-i \hl_1 \cdot (\xi_1 \bar{n}_i)} \int_{-\infty}^{\infty} d\xi_2 \, e^{-i \hl_2 \cdot (\xi_2 \bar{n}_i)} \langle k_i | \phi(\xi_1 \bar{n}_i) \phi(\xi_2 \bar{n}_i) \bpsi(0) | 0 \rangle \, , \nonumber \\
J^{fff}(k_i - \hl_1 - \hl_2, \hl_1, \hl_2)  &= \int_{-\infty}^{\infty} d\xi_1 \, e^{-i \hl_1 \cdot (\xi_1 \bar{n}_i)} \int_{-\infty}^{\infty} d\xi_2 \, e^{-i \hl_2 \cdot (\xi_2 \bar{n}_i)} \langle k_i | \bpsi(\xi_1 \bar{n}_i) \psi(\xi_2 \bar{n}_i) \bpsi(0) | 0 \rangle \, , \nonumber \\
J^{f\softs}(k_i + l , l ) &= \int d^4 y \, e^{i l \cdot y} \langle k_i | \, \frac{\delta S_I}{\delta \phi(y)} \,  \bpsi(0) | 0 \rangle \, , \nonumber \\
J^{f\partial s}(k_i - \hl, \hl) &= \int_{-\infty}^{\infty} d\xi \, e^{-i \hl \cdot (\xi \bar{n}_i)} \langle k_i | (\partial_T \phi )(\xi \bar{n}_i) \bpsi (0) | 0 \rangle \, .
\end{align}
where in the above $S_I$ is the interacting part of the action. The label $s$ in the superscripts above refers to a collinear scalar while the label $\underline{s}$ refers to a soft scalar. It will be convenient to refer to the jet components of reduced diagrams and to the jet functions themselves as leading jets, $fs$-jets, $fss$-jets, etc.\ depending on the jet's outgoing collinear and soft particles. The derivative label, $\partial$, indicates that we are expanding the hard part in the transverse component of the loop momentum of the particle whose label follows the $\partial$. In the $f\partial s$ jet example here, we are expanding the hard part in the transvere momentum of the collinear scalar, yielding a jet of $O(\lambda^2)$. The jet functions have one free Dirac index for each label $f$. This Dirac index is to be contracted with a matching index in the corresponding hard part. It is clear that the above sample definitions can be generalized to an arbitrary number of collinear or soft fermions and scalars. The hat over the collinear loop momentum arguments of the jet functions indicates that we are only retaining the part of the loop momentum collinear to the external momentum $k_i$, which is defined as $\hat{l} = l^+ n_i$.

The factorized form of the elastic amplitude, including terms of order up to $O(\lambda^2)$, was derived in \cite{Gervais:2017yxv} and is given by
\begin{align} \label{eq:elastic_factorized}
\mathcal{M}_{el} = & \left(\prod_{i=1}^n \tensor{J}{_i^f}\right)\otimes  H  \nonumber \\
&+ \sum_{i=1}^n \left(\prod_{j \neq i} \tensor{J}{_j^f}\right) \tensor{J}{_i^f^s} \otimes \tensor{H}{_i^f^s}+ \sum_{i=1}^n \left(\prod_{j \neq i} \tensor{J}{_j^f}\right) \tensor{J}{_i^f^\partial^s} \otimes \tensor{H}{_i^f^\partial^s} \nonumber \\
&+ \sum_{i=1}^n \left(\prod_{j \neq i} \tensor{J}{_j^f}\right) \tensor{J}{_i^f^s^s} \otimes \tensor{H}{_i^f^s^s} + \sum_{i=1}^n \left(\prod_{j \neq i} \tensor{J}{_j^f}\right) \tensor{J}{_i^f^f^f} \otimes \tensor{H}{_i^f^f^f} \nonumber \\
&+ \sum_{1\leq i < j \leq n} \left( \prod_{l\neq i,j} \tensor{J}{_l^f} \right) \tensor{J}{_i^f^s} \tensor{J}{_j^f^s} \otimes H_{ij}^{fs;fs} + \sum_{1\leq i < j \leq n} \left( \prod_{l\neq i,j} \tensor{J}{_l^f} \right) \tensor{J}{_i^f^\softs} \tensor{J}{_j^f^\softs} \, S_{ij} \otimes  H_{ij}^{f\softs; f\softs}  \nonumber \\
& + O (\lambda^3) \, .
\end{align}
The momenta arguments are left implicit but can be recovered from the definitions of the jet functions as in \eqref{eq:jet_def_examples}. The tensor product symbol $\otimes$ indicates that there is a contraction between the spinor indices of the jet functions and the corresponding indices carried by the hard part. It is also convenient to let the $\otimes$ symbol denote the convolution product of the hard part and the jet functions. This convolution is taken in the collinear components of the momenta of the particles associated with the fields appearing in \eqref{eq:jet_def_examples} -- see \cite{Gervais:2017yxv} for details.

In \eqref{eq:elastic_factorized}, the soft function $S_{ij}$ is a two-point function with soft external scalar momenta connecting to jets $i$ and $j$. It is useful to extend this notation in a straightforward way, so that for example, $S_{iikl}$ stands for a four-point function with soft external scalar momenta, two of which connect to jet $i$, and with the two others connecting to jets $k$ and $l$. In the notation for the hard parts, the superscripts indicate which non-leading jets are present, separated by semicolons. The subscripts indicate which external particle is coming from the corresponding non-leading jet in the superscript. For example, the hard part $H^{fff;fs}_{ij}$ has an $fff$-jet at the $i^{th}$ external particle and an $fs$-jet at the $j^{th}$ external particle.

In the interest of space,  we will not write the $O(\lambda^3)$ and $O(\lambda^4)$ terms in equation format. The $O(\lambda^3)$ terms derive from the diagrams shown in Figs.\   \ref{fig:g_3_partition_3}, \ref{fig:g_3_partition_21}, and \ref{fig:g_3_partition_111}. The associated factorized terms are listed in Table \ref{tab:gamma_3_factorized}. The $O(\lambda^4)$ terms correspond to the reduced diagrams shown in Figs.\ \ref{fig:g_4_partition_4},  \ref{fig:g_4_partition_31}, \ref{fig:g_4_partition_22}, \ref{fig:g_4_partition_211}, and \ref{fig:g_4_partition_1111}. The associated factors are displayed in Table \ref{tab:gamma_4_factorized}.  Finally, we also need to consider the ``exceptional contributions'' whose diagrams appear in Fig.\ \ref{fig:hard_connection_example}, with the associated factorized contributions listed in Table \ref{tab:exceptional_contributions}.  To generate the explicit $O(\lambda^3)$ and $O(\lambda^4)$ terms, one would need to contract the factors found in these tables analogously to \eqref{eq:elastic_factorized}.

It should be mentioned that additional jet functions need to be defined at $O(\lambda^2)$, $O(\lambda^3)$, and $O(\lambda^4)$. These arise when one expands the hard parts of lower order terms in the transverse components of the loop momenta of the collinear particles connecting hard parts to jets. We will omit these constructions here. Details can be found in \cite{Gervais:2017yxv}.


\begin{table}[h!]
\caption{The factors corresponding to the diagrams with $\gamma = 3$. A horizontal line separates factorized forms corresponding to distinct partitions of $\gamma = 3$. These factorized contributions are associated with the diagrams shown in Figs.\ \ref{fig:g_3_partition_3}, \ref{fig:g_3_partition_21}, and \ref{fig:g_3_partition_111}. }

\begin{center}
	\begin{tabular}{lcl} 
	\hline
	\hline
	Non leading jets                                                         &    Soft function              &     Hard part                \\ \hline
	   $J^{fsss}_i$                                                           &               1                      &          $  H_i^{fsss}$              \\
	   $J^{fffs}_i$                                                           &               1                      &          $ H_i^{fffs}   $            \\ \hline
	   $J^{fss}_i J^{fs}_j$                                             &             1                         &           $ H_{ij}^{fss;fs}$              \\
	   $J^{fff}_i J^{fs}_j$                                              &             $1$                    &          $ H_{ij}^{fff;fs} $              \\
	   $J^{fs\softs}_i J^{f\softs}_j$                              &             $S_{ij}$             &          $ H_{i}^{fs}$               \\ \hline
	   $J^{f\softs}_i J^{f\softs}_j J^{fs}_k$                &     $S_{ij}$                     &         $H^{fs}_{k}$                                  \\
	   $J^{fs}_i J^{fs}_j J^{fs}_k $                               &       1                              &         $H^{fs;fs;fs}_{ijk}$                  \\

\hline \hline
	\end{tabular}

\label{tab:gamma_3_factorized}
\end{center}
\end{table}



\begin{table}[h!]
\caption{The factors corresponding to the diagrams with $\gamma = 4$. A horizontal line separates factorized forms corresponding to distinct partitions of $\gamma = 4$. These factorized contributions are associated with the diagrams shown in Figs.\  \ref{fig:g_4_partition_4},  \ref{fig:g_4_partition_31}, \ref{fig:g_4_partition_22}, \ref{fig:g_4_partition_211}, and \ref{fig:g_4_partition_1111}.}

\begin{center}
	\begin{tabular}{lcl} 
	\hline
	\hline
	Non leading jets                                                         &    Soft function              &     Hard part                \\ \hline
	   $J^{fffss}_i$                                                          &               1                      &          $  H_i^{fffss}$              \\
	   $J^{fssss}_i$                                                          &               1                      &          $ H_i^{fssss}   $            \\ \hline
	   $J^{fss\softs}_i J^{f\softs}_j$                              &             $S_{ij}$            &           $ H_{ij}^{fss;f}$              \\
	   $J^{fff\softs}_i J^{f\softs}_j$                              &             $S_{ij}$             &          $ H_{ij}^{fff;f} $              \\
	   $J^{fffs}_i J^{fs}_j$                                              &             1                        &          $ H_{ij}^{fffs;fs}$               \\
	   $J^{f \softs \softs\softs}_i J^{f\softs}_j$            &              $S_{iiij}$            &          $H $                                  \\
	   $J^{fsss}_i J^{fs}_j$                                             &              $S_{ij}$             &         $ H_{ij}^{fsss;fs} $               \\ \hline
	   $J^{fss}_i J^{fss}_j$                                             &              1                        &          $H_{ij}^{fss;fss}$                 \\
	   $J^{fff}_i J^{fff}_j$                                             &               1                        &          $H_{ij}^{fff;fff}$                \\
	   $J^{fff}_i J^{fss}_j$                                             &               1                        &          $H_{ij}^{fff;fss}$                \\
	   $J^{fss}_i J^{fs\softs}_j$                                     &               $S_{ij}$             &          $H_{ij}^{fs;fs}$                \\
	   $J^{f\softs \softs}_i J^{f\softs \softs}_j$            &               $S_{iijj}$           &          $H$                                     \\ \hline
	   $J^{fss}_i J^{fs}_j J^{fs}_k$                               &                 1                      &           $H^{fss;fs;fs}_{ijk}$                \\
	   $J^{fss}_i J^{f\softs}_j J^{f\softs}_k$               &                 $S_{jk}$           &          $H^{fss}_{i}$               \\
	   $J^{fs\softs}_i J^{f\softs}_j J^{fs}_k$               &                 $S_{ij}$             &         $H^{fs;fs}_{ik}$               \\
	   $J^{f\softs \softs}_i J^{f\softs}_j J^{f\softs}_k$   &            $S_{iijk}$           &         $H$                                     \\
	   $J^{fff}_i J^{fs}_j J^{fs}_k$                              &                  1                       &          $H^{fff;fs;fs}_{ijk}$                 \\
	   $J^{fff}_i J^{f\softs}_j J^{f\softs}_k$               &                 $S_{jk}$             &        $H^{fff}_{i}$                   \\
	   $J^{f\softs}_i J^{f\softs}_j J^{f\softs}_k J^{f\softs}_l$  &     $S_{ijkl}$        &         $H$                                  \\
	   $J^{f\softs}_i J^{f\softs}_j J^{fs}_k J^{fs}_l$   &                 $S_{ij}$             &         $H^{fs;fs}_{kl}$                  \\
	   $J^{fs}_i J^{fs}_j J^{fs}_k J^{fs}_l$                  &                   1                       &         $H^{fs;fs;fs;fs}_{ijkl}$                  \\

\hline \hline
	\end{tabular}

\label{tab:gamma_4_factorized}
\end{center}
\end{table}



\begin{table}[h!]
\caption{The exceptional factorized terms contributing to the elastic amplitude. In the soft function subscript, the label $H$ indicates that a soft scalar is attaching the soft cloud to the hard part. Likewise, in the hard part superscript, the label $S$ indicates that a soft scalar connects the hard part to the soft cloud. These factorized contributions are associated with the diagrams shown in Fig.\ \ref{fig:hard_connection_example}.}

\begin{center}
	\begin{tabular}{clcl} 
	\hline
	\hline
	Order	                               &         Non leading jets                    &    Soft function              &     Hard part                \\ \hline
       $\lambda^3$               &      $J^{f\softs}_i$                       &               $S_{iH}$         &          $  H^{S}$              \\
       $\lambda^4$               &      $J^{f\softs}_i J^{fs}_j$        &               $S_{iH}$         &          $ H_j^{fs;S}   $            \\ 
       $\lambda^4$               &      $J^{fs\softs}_i $                    &            $ S_{iH}$           &           $ H_{i}^{fs;S}$              \\

\hline \hline
	\end{tabular}

\label{tab:exceptional_contributions}
\end{center}
\end{table}



\section{Graviton Emission} \label{sec:graviton_emission}

\subsection{Modification of Power Counting from Graviton Emission} \label{subsec:graviton_emission_power_counting}

The amplitude for soft graviton emission from a jet is generated by inserting a matter-graviton vertex into the matrix element definition of the corresponding jet function. For instance, in the case of a leading jet, we obtain the following radiative jet function,
\begin{align}
J_{\mu\nu}(k_i,q) &=\int d^4 x\,e^{-iq\cdot x} \left \langle k_i \left | \mathcal{T}^* \,( iT_{\mu\nu}(x) \phi_i(0)) \right| 0 \right\rangle\,,
\end{align}
where $T_{\mu\nu}$ is the stress tensor through which we assume gravity couples to matter. This stress tensor is defined by taking the variational derivative of the matter action, as shown in the following equation \cite{BjerrumBohr:2004mz, Veltman:1975vx},
\begin{align}
\delta S_{matter} \equiv \frac{1}{2} \int d^4x \,  T^{\mu\nu}(x) \, \delta g_{\mu\nu}(x).
\end{align}
Our prescription for coupling matter to gravity is to make insertions of the operator $iT_{\mu\nu}$ to generate a matter graviton vertex. The dynamics of the graviton field is dictated by the full Einstein-Hilbert action. We will, however, not be concerned with graviton loops in this paper.

We now analyze the effect of emitting a graviton from one of the lines or vertices in the reduced diagrams that give the term of order $O(\lambda^0)$ up to $O(\lambda^4)$ identified in Sec.\ \ref{sec:elastic_amplitude_expansion}. In doing this, we need to take into account the possibility of emitting the graviton from a collinear line, a soft line, or directly from the vertices of Yukawa and scalar theory.

\begin{figure}[h!]
\centering
\includegraphics[width = 0.40 \textwidth]{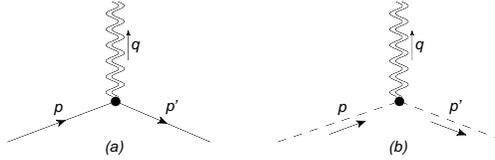}
\caption{The fermion-graviton (ffG) and scalar-graviton (ssG) vertices.}
\label{fig:matter_graviton_vertices}
\end{figure}

We begin by considering graviton emission from a fermion line. The graviton-fermion vertex is given by
\begin{align}
V_{ffG}^{\mu\nu} &= \frac{-i \kappa}{2} \left[\frac{1}{4}(\gamma^\mu (p+p^\prime)^\nu + \gamma^\nu ( p+ p^\prime)^\mu -  \eta^{\mu\nu} \left( \frac{1}{2} ( \pslash +  \pslash^\prime) - m\right)\right] \, ,
\end{align}
where $\kappa$ is Newton's constant and $p$ and $p^\prime$ are the incoming and outgoing fermion momenta -- see Fig.\ \ref{fig:matter_graviton_vertices} (a).


Starting from a fermion line carrying momentum $p$, emitting a graviton from this line changes the propagator according to
\begin{align}\label{eq:graviton_fermion_emission}
\frac{i}{\pslash - m} \mapsto \frac{i \kappa}{8} \left[ \frac{\pslash + m}{p^2 - m^2} (\gamma^\mu ( 2 p+ q)^\nu + \gamma^\nu ( 2p + q)^\mu) \frac{\pslash + \qslash + m}{(p+q)^2 - m^2} - \frac{2 \eta^{\mu\nu}}{\pslash + \qslash - m} - \frac{2 \eta^{\mu\nu}}{\pslash - m} \right] \, .
\end{align}

Suppose we had started with a collinear fermion line. On the right hand side of the arrow, the leading term is the first on the left. It has two collinear denominators, each of which scales as $\lambda^2$, and a numerator whose leading term scales as $\lambda^0$ when $\mu\nu = ++$. Since we started with only one collinear denominator scaling as $\lambda^2$, we conclude that the net effect of adding a graviton is to reduce the degree of divergence $\gamma$ of the whole diagram by $2$.

On the other hand, suppose we had started with a soft fermion. Then we begin with a propagator where the numerator and denominator are dominated by the mass term. Therefore, this soft fermion propagator scales as $\lambda^{-1}$ and, as seen from \eqref{eq:graviton_fermion_emission}, emitting the graviton from this line either introduces an additional soft fermion propagator and a numerator that scales as $\lambda^2$, or does not introduce any new factor as in the rightmost two terms. The net effect is therefore to leave the scaling power $\gamma$ unchanged.

Consider now emitting the soft graviton from a scalar line as in Fig.\ \ref{fig:matter_graviton_vertices} (b). The scalar-graviton vertex is given by
\begin{align}
V_{ssG}^{\mu\nu} &= \frac{-i\kappa}{2} ( p^\mu p^{\prime \nu} + p^\nu p^{\prime \mu} - \eta^{\mu\nu}(p \cdot p^\prime - m^2))\, .
\end{align}
The starting massless scalar propagator is therefore replaced according to
\begin{align}
\frac{i}{p^2} \mapsto \frac{i\kappa}{2} \frac{1}{p^2} \frac{1}{(p+q)^2} (2 p^\mu p^\nu + p^\mu q^\nu + p^\nu q^\mu - \eta^{\mu\nu}(p^2 + p\cdot q)) \, .
\end{align}
If the original scalar is collinear, then as before we have added a denominator scaling as $\lambda^2$ and a numerator scaling as $1$ when $\mu\nu = ++$, for a net change of $\gamma \mapsto \gamma -2$. If, however, the scalar is soft, then the addition of a denominator scaling as $\lambda^4$ is entirely compensated by the numerator scaling as $\lambda^4$ as well. Hence, emitting a soft graviton from a soft scalar does not alter the degree of divergence of the whole diagram.

\begin{figure}[h!]
\centering
\includegraphics[width = 0.40 \textwidth]{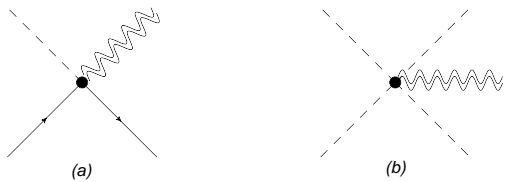}
\caption{The scalar-two-fermion-graviton (sffG) and the four-scalar-graviton (ssssG) vertices.}
\label{fig:potential_graviton_vertices}
\end{figure}

It remains to analyze the effect of emitting a graviton directly from the interaction vertices of Yukawa and scalar theory -- see Fig.\ \ref{fig:potential_graviton_vertices}. Let $g$ be the Yukawa coupling and $g^\prime$ be the four-scalar coupling. Then emitting the graviton from the Yukawa vertex introduces a factor of $-\frac{ig\kappa}{2} \eta^{\mu\nu}$ into the diagram and nothing else. This does not alter the degree of divergence. Similarly, emitting the soft graviton from the four-scalar vertex introduces a factor of $-\frac{ig^\prime \kappa}{2} \eta^{\mu\nu}$ and does not alter the degree of divergence.

Finally, we observe that a soft graviton insertion onto a hard line will not affect its scaling. Therefore, emitting a soft graviton from the hard part of a diagram has no effect on the degree of divergence.


The effects of emitting a soft graviton from a reduced diagram of the elastic amplitude are gathered in Table \ref{tab:gamma_graviton_effect}. We see that the greatest enhancement occurs when the graviton is emitted from a collinear fermion or scalar line. Further, in this case, the infrared degree of divergence of the newly obtained radiative diagram is equal to the degree of divergence of the original elastic amplitude diagram minus $2$.


\begin{table}[h!]
\caption{This table describes how to obtain the degree of divergence of a radiative diagram from the degree of divergence of the corresponding elastic diagram. The effect of graviton emission depends on which component of the elastic diagram the graviton is emitted from. }

\begin{center}
	\begin{tabular}{cc} 
	\hline
	\hline
	Component emitting the soft graviton       &         Net effect on degree of divergence of elastic diagram     \\ \hline
Collinear fermion line  &            -2         \\
Collinear scalar line    &             -2         \\
Soft fermion line         &             +0         \\
Soft scalar line            &             +0         \\
Yukawa vertex            &             +0         \\
Four-scalar vertex      &              +0        \\
Hard part                    &              +0         \\

\hline \hline
	\end{tabular}

\label{tab:gamma_graviton_effect}
\end{center}
\end{table}


In Eq.\ \eqref{eq:soft_theorem_lambda_expansion}, we noted that the soft graviton theorem can be thought of as an expansion of the radiative amplitude in power of the small scale $\lambda$. Any diagram contributing to this expansion can be derived by attaching a graviton vertex to a diagram contributing to the elastic amplitude. Further, we have just shown that the resulting radiative diagram will be of order $O(\lambda^{\gamma - 2})$, where we assume that the original elastic amplitude diagram was of order $O(\lambda^\gamma)$. Therefore, to get all contributions to the radiative amplitude between $O(\lambda^{-2})$ and $O(\lambda^2)$, we need to begin our construction with all elastic diagrams of orders between $O(\lambda^0)$ and $O(\lambda^4)$, as we claimed in Sec.\ \ref{sec:elastic_amplitude_expansion}.

Graviton emission from the hard part or the soft cloud needs to be considered only for elastic diagrams with scaling power $\gamma$ between $0$ and $2$. Indeed, emitting a graviton from the hard part or the soft cloud has no effect on the degree of divergence, and hence we need to start from an elastic diagram with $\gamma \leq 2$ to remain within the order of accuracy of the soft graviton theorem.

\subsection{Adapting Low's Analysis to Factorized Amplitudes} \label{subsec:low_analysis}

Our goal is to extend Low's analysis to gravity for a factorized amplitude. Since the pinch surfaces of the radiative amplitude are in correspondence with the reduced diagrams of the elastic amplitude, it will be enough to show how this is done by considering the reduced diagram with $n$ external leading jets. The extension of Low's analysis to diagrams containing non-leading jets involves repeated use of the off-shell Ward identity discussed in Appendix \ref{sec:off_shell_ward}, and follows the lines of the argument we are about to present. The analysis here is similar to Ref.\ \cite{Bern:2014vva}, but as in Ref.\ \cite{Gervais:2017yxv}, we are going to give details on the role of kinematics.

We emphasize the necessity of using the construction of Burnett and Kroll \cite{Burnett:1967km} to transition between the radiative and elastic kinematics. This point has mostly been overlooked in the litterature on the soft graviton theorem \cite{Gervais:2017yxv}. The issue resides in the expansion of the elastic amplitude in powers of $q$,
\begin{align}\label{eq:conventional_elastic_expansion}
\mathcal{M}_{el} & (k_1, \dots, k_i + q, \dots,  k_n) = \nonumber \\
&  \mathcal{M}_{el}(k_1, \dots, k_n) + q^\alpha \frac{\partial \mathcal{M}_{el}}{\partial k_i^\alpha} (k_1, \dots, k_n) + \frac{1}{2} q^\alpha q^\beta \frac{\partial^2 \mathcal{M}_{el}}{\partial k_i^\alpha \partial k_i^\beta} (k_1, \dots, k_n) + O(q^3) \, .
\end{align}
The momenta $k_1, \dots, k_n$ satisfy momentum conservation in the form
\begin{align}
k_1 + \dots+ k_n = -q \, ,
\end{align}
assuming all external momenta are outgoing. The elastic amplitude $\mathcal{M}_{el}$ in \eqref{eq:conventional_elastic_expansion}, however,  is defined on the locus of momenta $k_1^\prime, \dots, k_n^\prime$ satisfying the constraint
\begin{align} \label{eq:elastic_momentum_conservation}
k_1^\prime + \dots + k_n^\prime = 0 \, .
\end{align}
It thus appears that \eqref{eq:conventional_elastic_expansion} is ill defined. The solution, following \cite{Gervais:2017yxv, Burnett:1967km}, is to define a set of elastic momenta $k_1^\prime, \dots, k_n^\prime$ that are shifted from the radiative configuration $k_1, \dots, k_n$ by deviation vectors $\xi_i(q)$ according to
\begin{align} \label{eq:real_def_xi}
k_i = k_i^\prime + \xi_i\, \text{ for $i = 1, \dots, n$} \, .
\end{align} 
Momentum conservation imposes the requirement that
\begin{align} \label{eq:definition_xi}
\sum_{i = 1}^{n} \xi_i = - q \, .
\end{align}
Additionally, we demand that $\xi_i(q) = O(q)$ for all $i =1, \dots, n$ and that the $k_i^\prime$ be on-shell. Ref.\ \cite{Gervais:2017yxv} shows explicitly how to construct the $\xi_i$'s to $O(q)$. The extension of this construction to $O(q^2)$ is straightforward.

The crux of Low's argument is the use of the on-shell gravitational Ward identity \cite{Weinberg:1964ew},
\begin{align}\label{eq:GravWard}
q^\mu \mathcal{M}_{\mu\nu}&=q^\mu (\mathcal{M}^{ext}_{\mu\nu}+\mathcal{M}^{int}_{\mu\nu})=0 \, ,
\end{align}
 to relate the external emission amplitude to the internal amplitude. In the context of factorized diagrams, the external amplitude is defined as the amplitude to emit a graviton from a jet, whether leading or non-leading. The internal emission, on the other hand, consists in the amplitude to emit the graviton from the hard part, or the soft cloud, if it is present in the diagram at hand. It is important to realize that the Ward identity holds separately for each reduced diagram -- see Appendix \ref{sec:off_shell_ward}. This property of the Ward identity ensures that it is legitimate to consider individually each of the factorized amplitudes identified in Sec.\ \ref{subsec:elastic_factorization}.

We will also use an off-shell Ward identity that applies to jet functions -- see Appendix \ref{sec:off_shell_ward}. The result is spin-dependent in the external line, and is given for leading jets by \cite{Brout:1966oea,Bessler:1969py,Coriano:2011zk,Coriano:2012cr}
\begin{align}\label{eq:JetWard}
q^\mu J^s_{\mu\nu}(k_i,q) &= \ J^s (k_i)\, k_{i,\nu} & \text{for scalars, and} \nonumber \\
q^\mu J^f_{\mu\nu}(k_i,q) &= \ J^f (k_i)\, \left( k_{i,\nu} - \frac{1}{2}q^\mu \sig\right) & \text{for fermions,}
\end{align}
where the spinor indices of the matrix $\sig = \frac{1}{4}[\gamma_\mu,\gamma_\nu]$ are summed with those of the $i^{th}$ jet function $J^f(k_i)$ and the corresponding indices in the hard part. This form is valid for the case of an outgoing graviton and requires an overall minus sign for an incoming graviton.

\begin{figure}[h!]
\centering
\includegraphics[width = 0.75\textwidth]{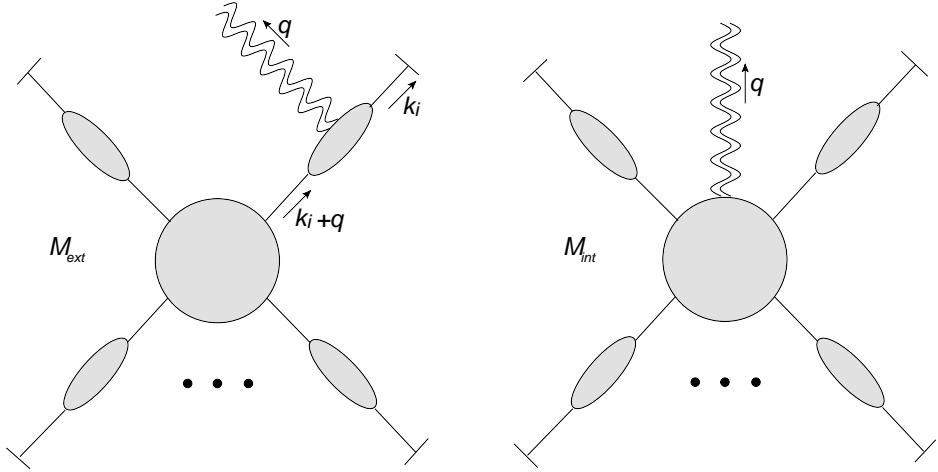}
\caption{The radiated graviton can be emitted either from an outgoing jet or the internal hard function.}
\label{fig:ext_and_int}
\end{figure}

We begin with the case of external scalars. In the case of diagrams with leading jets only, the basic factorization we use as a starting point is
\begin{align} \label{eq:factorized_graviton_emission}
\mathcal{M}_{el} &= \left( \prod_{i=1}^n J^s(k_i) \right) \otimes H(k_1, \dots, k_n) \, , \nonumber \\
\mathcal{M}_{ext}^{\mu\nu} &= \sum_{i=1}^n \left( \prod_{j \neq i} J^s(k_j) \right) J^{s,\mu\nu} (k_i, q) \otimes H(k_1, \dots, k_i + q, \dots, k_n) \, , \nonumber \\
\mathcal{M}_{int}^{\mu\nu} &= \left( \prod_{i=1}^n J^s(k_i) \right) \otimes H^{\mu\nu} (k_1, \dots, k_n, q) \, .
\end{align}
This factorization is illustrated in Fig.\ \ref{fig:ext_and_int}. Since both  $H(k_1 \dots k_i+q \dots k_n)$ and $H_{\mu\nu}(k_1\dots  k_n,q)$ are fully infrared finite, we can safely expand them in powers of $q$ at fixed values of the $k_i$, so long as the internal lines are off-shell by a scale set by the invariants formed by the $k_i\cdot k_j$, $i\ne j$, and all loop integrals converge independently of $q$.   We note that this condition fails in the external jet subdiagrams in general, where, when loop momenta become collinear to $k_i$, we cannot expand around $q^\mu=0$.  This was noted by Del Duca \cite{DelDuca:1990gz} in the context of Low's theorem.   Once the jets are factored, however, the remaining hard subdiagrams can be expanded in powers of $q$ since the soft graviton insertion will not alter the power behavior of their loop integrals -- see the discussion in Sec.\ \ref{subsec:graviton_emission_power_counting}.


The objective is to solve the two constraints $q_\mu \mathcal{M}_{int}^{\mu\nu} = - q_\mu \mathcal{M}_{ext}^{\mu\nu}$ and  $q_\nu \mathcal{M}_{int}^{\mu\nu} = - q_\nu \mathcal{M}_{ext}^{\mu\nu}$ for the radiative hard part $H^{\mu\nu}$. Our approach is motivated by the elementary methods for solving linear equations. Namely, we first find a particular solution $H^{\mu\nu}$ that solves both constraints simultaneously. This will be done by solving one constraint and demanding that $H^{\mu\nu}$ be symmetric. Any other simultaneous solution to the two constraints must differ from $H^{\mu\nu}$ by a gauge invariant quantity $B^{\mu\nu}$ obeying $q_\mu B^{\mu\nu} = 0$ and $q_\nu B^{\mu\nu} = 0$. We will then show that, under the condition that the hard part may not contain any singularity in $q$, such a $B^{\mu\nu}$ must vanish to $O(q)$.

We now write out each term in the Ward identity \eqref{eq:GravWard} explicitly using \eqref{eq:JetWard},
\begin{align}
q_\mu \mathcal{M}_{ext}^{\mu\nu} &= \left( \prod_{j=1}^n J^s(k_j) \right) \otimes \, \sum_{i=1}^n    k_i^\nu H(k_1, \dots, k_i + q, \dots, k_n) \, , \nonumber \\
q_\mu \mathcal{M}_{int}^{\mu\nu} &= \left( \prod_{j=1}^n J^s(k_j) \right) \otimes q_\mu  H^{\mu\nu} (k_1, \dots, k_n, q) \, ,
\end{align}
and see that
\begin{align}\label{eq:ward_identity_explicit}
q_\mu H^{\mu\nu} ( k_1, \dots, k_n, q) &= - \sum_{i=1}^n k_i^\nu H(k_1, \dots, k_i + q, \dots, k_n) \, ,
\end{align}
with $H$ the hard function of the elastic amplitude as in \eqref{eq:factorized_graviton_emission}.

Next, we may proceed with the expansion about the elastic configuration following \eqref{eq:real_def_xi} and \eqref{eq:definition_xi}. Consequently, the hard part is no longer expanded only in $q$, but also in the $\xi_i$'s,
\begin{align} \label{eq:expansion_hard_q_xi}
H(k_1, \dots, k_i + q, \dots, k_n) =& \,  H(k_1^\prime + \xi_1, \dots, k_i^\prime + \xi_i + q, \dots, k_n^\prime + \xi_n) \nonumber \\
=&  \, H(k_1^\prime, \dots, k_n^\prime) + q^\rho \frac{\partial H}{\partial k_i^{\prime\rho}} + \sum_{j=1}^n \xi_j^\rho \frac{\partial H}{\partial k_j^{\prime\rho}} \nonumber \\
&+ \frac{1}{2} q^\rho q^\sigma \frac{\partial^2 H}{\partial k_i^{\prime\rho} \partial k_i^{\prime\sigma}} + \sum_{j=1}^n q^\rho \xi_j^\sigma \frac{\partial^2 H}{\partial k_i^{\prime\rho} \partial k_j^{\prime\sigma}} + \frac{1}{2} \sum_{j,l =1}^n \xi_j^\rho \xi_l^\sigma \frac{\partial^2 H}{\partial k_j^{\prime\rho} \partial k_l^{\prime\sigma}} \, .
\end{align}
Substituting this into \eqref{eq:ward_identity_explicit}, we find
\begin{align} \label{eq:contracted_hard_part}
q_\mu H^{\mu\nu}(k_1, \dots, k_n, q)
= & \,  q^\nu \left( H + \sum_{j=1}^n \xi_j \cdot \frac{\partial H}{\partial k_j^\prime} \right) \nonumber \\
& \, - \sum_{i=1}^n \left( k_i^{\prime\nu} q \cdot \frac{\partial H}{\partial k_i^\prime} \right) \nonumber \\
& \, -\sum_{i=1}^n \left(\xi_i^\nu q \cdot \frac{\partial H}{\partial k_i^\prime} + \frac{1}{2} k_i^{\prime\nu} (qq) \cdot \frac{\partial^2H}{\partial k_i^\prime \partial k_i^\prime} + \sum_{j=1}^n k_i^{\prime \nu} (q \xi_j) \cdot \frac{\partial^2 H}{\partial k_i^\prime \partial k_j^\prime} \right) \, ,
\end{align} 
where we have used the kinematics relations \eqref{eq:elastic_momentum_conservation} to \eqref{eq:definition_xi}. The ``$\cdot$'' notation stands for a contraction between matching Minkowski indices, with the understanding that $q$ is always contracted with $\frac{\partial}{\partial k_i}$ and $\xi_j$ is always paired with $\frac{\partial}{\partial k_j}$. For example, $(q \xi_j ) \cdot \frac{\partial H}{\partial k_i \partial k_j} = q^\alpha \xi_j^\beta \cdot \frac{\partial H}{\partial k_i^\alpha \partial k_j^\beta}$. Factoring out $q_\mu$ from \eqref{eq:contracted_hard_part}, we obtain a solution to the Ward identity \eqref{eq:contracted_hard_part},
\begin{align} \label{eq:factored_hard_part}
\tilde{H}^{\mu\nu} &= \eta^{\mu\nu} \left( H + \sum_{j=1}^n \xi_j \cdot \frac{\partial H}{\partial k_j^\prime}  \right) \nonumber \\
& \, - \sum_{i=1}^n \left( k_i^{\prime \nu} \frac{\partial H}{\partial k_{i\mu}^\prime} \right) \nonumber \\
& \, - \sum_{i=1}^n \left(\xi_i^\nu \frac{\partial H}{\partial k_{i\mu}^\prime} + \frac{1}{2} k_i^{\prime\nu} q \cdot \frac{\partial^2 H}{\partial k_{i\mu}^\prime \partial k_i^\prime} + \sum_{j=1}^n k_i^{\prime \nu} \xi_j \cdot  \frac{\partial^2 H}{\partial k_{i\mu}^\prime \partial k_j^\prime }\right)\, .
\end{align}
It is not yet clear, however, whether $\tilde{H}^{\mu\nu}$ gives the full radiative hard part. We could, for instance, be missing a separately gauge invariant contribution. Further, the radiative hard part must be symmetric under interchange of $\mu$ and $\nu$, which is not immediately obvious from \eqref{eq:factored_hard_part}. We will return to the issue of separately gauge invariant contributions below and focus on the question of symmetry first.

For scalars, angular momentum conservation implies
\begin{align} \label{eq:scalar_angular_momentum_conservation}
\sum_{i=1}^n \mathcal{J}_{i,\alpha\beta}  H = \sum_{i=1}^n \left( k_{i\alpha}^{\prime} \frac{\partial}{\partial k_i^{\prime \beta}} - k^\prime_{i\beta} \frac{\partial}{\partial k_i^{\prime \alpha}}\right) H &= 0 \, ,
\end{align}
so that the second term in \eqref{eq:factored_hard_part} is actually symmetric under interchange of $\mu$ and $\nu$. The operator $\mathcal{J}_i$ is the angular momentum operator from \eqref{eq:orbital}. The clearest symmetrization of the $\xi_i$-independent $O(q)$ term is
\begin{align}
\frac{1}{2} k_i^{\prime \nu} q \cdot \frac{\partial^2 H}{\partial k_{i\mu}^\prime \partial k_i^\prime} + \frac{1}{2} k_i^{\prime \mu} q \cdot \frac{\partial^2 H}{\partial k_{i\nu}^\prime \partial k_i^\prime} - \frac{1}{2} k_i^\prime \cdot q \frac{\partial^2 H}{\partial k_{i\mu}^\prime \partial k_{i\nu}^\prime} \, ,
\end{align}
where, when contracting with $q^\mu$, the second and third terms cancel and, hence, we recover \eqref{eq:contracted_hard_part}. We reiterate that we are enforcing the symmetry of $H^{\mu\nu}$ in order to construct a particular simultaneous solution to both $q_\mu \mathcal{M}_{int}^{\mu\nu} = - q_\mu \mathcal{M}_{ext}^{\mu\nu}$ and  $q_\nu \mathcal{M}_{int}^{\mu\nu} = - q_\nu \mathcal{M}_{ext}^{\mu\nu}$. The uniqueness of this solution will be shown at the end of this subsection. This leaves us with the $O(\lambda^2)$ $\xi_i$-dependent terms,
\begin{align}
- \sum_{i=1}^n \left( \xi_i^\nu \frac{\partial H}{\partial k_{i\mu}^\prime} + \sum_{j=1}^n k_i^{\prime \nu} \xi_j \cdot \frac{\partial^2 H}{\partial k_{i\mu}^\prime \partial k_j^\prime} \right) &= - \sum_{j=1}^n \xi_j \cdot \frac{\partial}{\partial k_j^{\prime}} \left( \sum_{i=1}^n  k_{i}^{\prime\nu} \frac{\partial H}{\partial k_{i\mu}^{\prime}}\right) \, ,
\end{align}
which can also be symmetrized using Eq. \eqref{eq:scalar_angular_momentum_conservation}.

In summary, we find that the extension of the soft graviton theorem including the $\xi_i$'s of Burnett and Kroll is 
\begin{align} \label{eq:radiative_hard_scalars}
H^{\mu\nu} &=  \eta^{\mu\nu} \left( H + \sum_{j=1}^n \xi_j \cdot \frac{\partial H}{\partial k_j^\prime}  \right)  - \sum_{i=1}^n \left( k_i^{\prime \nu} \frac{\partial H}{\partial k_{i\mu}^\prime} \right) \nonumber \\
& \, - \sum_{i=1}^n \left[\frac{1}{2} k_i^{\prime \nu} q \cdot \frac{\partial^2 H}{\partial k_{i\mu}^\prime \partial k_i^\prime} + \frac{1}{2} k_i^{\prime \mu} q \cdot \frac{\partial^2 H}{\partial k_{i\nu}^\prime \partial k_i^\prime} - \frac{1}{2} k_i^\prime \cdot q \frac{\partial^2 H}{\partial k_{i\mu}^\prime \partial k_{i\nu}^\prime} + \sum_{j=1}^n \xi_j \cdot \frac{\partial}{\partial k_j^{\prime}} \left( \sum_{i=1}^n  k_{i}^{\prime\nu} \frac{\partial H}{\partial k_{i\mu}^{\prime}}\right)\right]\, ,
\end{align}
where, again, \eqref{eq:scalar_angular_momentum_conservation} ensures the symmetry in $\mu$ and $\nu$.

In the case of fermions, the relevant jet Ward identity is given in \eqref{eq:JetWard},
\begin{align} \label{eq:jet_ward_fermion}
q_\mu J^{f,\mu\nu}(k_i,q) &= J^f (k_i)\left( k_i^\nu - \frac{1}{2} q_\mu \sigma_i^{\mu\nu} \right)\, .
\end{align}
Applying this Ward identity to the factorized form of the radiative amplitude with leading fermionic jets only, we find that compared with the scalar case, the radiative hard part contracted with $q_\mu$ has only the following additional terms,
\begin{align}
q_\mu \Delta H_{fer}^{\mu\nu} &=  \sum_{i=1}^n \frac{1}{2} q_\mu \sigma_i^{\mu\nu} \left( H + \sum_{j=1}^n \xi_j \cdot \frac{\partial H}{\partial k_j^\prime} + q \cdot \frac{\partial H}{\partial k_i^\prime} \right) \, ,
\end{align}
which are $\frac{1}{2} q_\mu \sigma_i^{\mu\nu}$ times the terms in \eqref{eq:expansion_hard_q_xi} of order up to $O(q)$.

Working first at $O(q)$ in the fermionic version of \eqref{eq:contracted_hard_part},
\begin{align} \label{eq:fermion_hard_q_zero}
(q_\mu H^{\mu\nu}) \big|_{O(q)} &=q^{\nu} H -  \sum_{i=1}^n \left( k_i^{\prime\nu} q \cdot \frac{\partial H}{\partial k_i^\prime}  -  \frac{1}{2} q_\mu \sigma_i^{\mu\nu} H \right) \, .
\end{align}
From angular momentum conservation, we now have the relation
\begin{align} \label{eq:fermion_angular_momentum_conservation}
\sum_{i=1}^n \mathcal{J}_i H = \sum_{i=1}^n \left( k_i^\alpha \frac{\partial}{\partial k_{i\beta}^\prime} - k_i^\beta \frac{\partial}{\partial k_{i\alpha}^\prime} + \sigma_i^{\alpha\beta} \right) H &= 0 \, ,
\end{align}
where $\mathcal{J}_i$ is the angular momentum operator from \eqref{eq:orbital}. This allows us to rewrite \eqref{eq:fermion_hard_q_zero} as 
\begin{align}
(q_\mu H^{\mu\nu}) \big|_{O(q)} &= q^\nu H - \frac{1}{2} \sum_{i=1}^n \left( k_i^{\prime\nu} q \cdot \frac{\partial H}{\partial k_i^\prime}  +  k_i^\prime \cdot q \frac{\partial H}{\partial k_{i\nu}^\prime} \right) \, .
\end{align}
Factoring out $q_\mu$ yields the symmetric combination
\begin{align}
H^{\mu\nu} \big|_{O(q^0)} &= \eta^{\mu\nu} H  - \frac{1}{2} \sum_{i=1}^n \left( k_i^{\prime\nu} \frac{\partial H}{\partial k_{i\mu}^\prime}  +  k_i^{\prime\mu} \frac{\partial H}{\partial k_{i\nu}^\prime} \right) \, ,
\end{align}
which is the same as in the scalar case \eqref{eq:radiative_hard_scalars}.

Consider next the $O(q^2)$ terms in $q_\mu H^{\mu\nu}$,
\begin{align}\label{eq:radiative_hard_fermions}
(q_\mu H^{\mu\nu}) \big|_{O(q^2)} = - \sum_{i=1}^n \bigg( & \xi_i^\nu q \cdot \frac{\partial H}{\partial k_i^\prime} + \frac{1}{2} k_i^{\prime \nu} (qq) \cdot \frac{\partial^2 H}{\partial k_i^\prime \partial k_i^\prime} + \sum_{j=1}^n k_i^{\prime\nu} (q\xi_j) \cdot \frac{\partial^2 H}{\partial k_i^\prime \partial k_j^\prime} \nonumber \\
& - \frac{1}{2} q_\mu \sigma_i^{\mu\nu} q \cdot \frac{\partial H}{\partial k_i^\prime}  - \frac{1}{2} q_\mu \sigma_i^{\mu\nu} \sum_{j=1}^n \xi_j \cdot \frac{\partial H}{\partial k_j^\prime} \bigg)  + q^\nu \sum_{j=1}^n \xi_j \cdot \frac{\partial H}{\partial k_j^\prime} \, .
\end{align}
Using angular momentum conservation, it is possible to reorganize the $\xi_i$-dependent terms using steps similar to those in the scalar case. The result is that the $O(q^2)$ terms above become
\begin{align}
 (q_\mu H^{\mu\nu}) \big|_{O(q^2)} = & \,  - \sum_{i=1}^n \left(  \frac{1}{2} k_i^{\prime \nu} (qq) \cdot \frac{\partial^2 H}{\partial k_i^\prime \partial k_i^\prime} - \frac{1}{2} q_\mu \sigma_i^{\mu\nu} q \cdot \frac{\partial H}{\partial k_i^\prime} \right) \nonumber \\
 &   -  \frac{1}{2} q_\mu \sum_{i,j =1}^n \xi_j \cdot \frac{\partial }{\partial k_j^\prime} \left(  k_i^{\prime \mu} \frac{\partial H}{\partial k_{i\nu}^{\prime}} + k_i^{\prime\nu} \frac{\partial H}{\partial k_{i\mu}^{\prime}} \right)  + q^\nu \sum_{j=1}^n \xi_j \cdot \frac{\partial H}{\partial k_j^\prime}\, .
\end{align}
Factoring out $q_\mu$ from this expression, we may finally write down the full radiative hard part when all external particles are fermions,
\begin{align}\label{eq:radiative_hard_part_fermions}
H^{\mu\nu} = & \, \eta^{\mu\nu} \left( H + \sum_{j=1}^n \xi_j \cdot \frac{\partial H}{\partial k_j^\prime} \right) \nonumber \\
&- \frac{1}{2} \sum_{i=1}^n \left( k_i^{\prime\mu} \frac{\partial H}{\partial k_{i\nu}^\prime} + k_i^{\prime\nu} \frac{\partial H}{ \partial k_{i\mu}^\prime} \right) \nonumber \\
& + \frac{1}{2} \sum_{i=1}^n \left(k_i^\prime \cdot q \frac{\partial^2 H}{\partial k_{ i \mu}^\prime \partial k_{i\nu}^\prime} - k_i^{\prime \mu} q^\rho \frac{\partial^2 H}{\partial k_i^{\prime \rho} \partial k_{i\nu}^\prime} - k_i^{\prime \nu} q^\rho \frac{\partial^2 H}{\partial k_i^{\prime \rho} \partial k_{i\mu}^\prime} + q_\rho \sigma_i^{\rho \mu} \frac{\partial H}{\partial k_{i\nu}^\prime} + q_\rho \sigma_i^{\rho \nu} \frac{\partial H}{\partial k_{i\mu}^\prime} \right) \nonumber \\
&   -  \frac{1}{2} \sum_{i,j=1}^n \xi_j \cdot \frac{\partial }{\partial k_j^\prime} \left( k_i^{\prime \mu} \frac{\partial H}{\partial k_{i\nu}^{\prime}} + k_i^{\prime\nu} \frac{\partial H}{\partial k_{i\mu}^{\prime}} \right) \nonumber \\
& + O(q^2) \, ,
\end{align}
where the $\xi_i$-dependent terms have been symmetrized using \eqref{eq:fermion_angular_momentum_conservation}. We will address the relation of this result to the CS formula \eqref{eq:CSresult}-\eqref{eq:orbital} in Sec.\ \ref{subsec:graviton_low_energy}.

In order to complete our derivation of the radiative hard parts \eqref{eq:radiative_hard_scalars} and \eqref{eq:radiative_hard_part_fermions}, we show that any supplementary contributions $B_{\mu\nu}$ to $H_{\mu\nu}$ statisfying $q^\mu B_{\mu\nu} =0$ and $q^\nu B_{\mu\nu} = 0$  must vanish to this order.  This is equivalent to showing the uniqueness of our solution $H^{\mu\nu}$ to the equations  $q_\mu \mathcal{M}_{int}^{\mu\nu} = - q_\mu \mathcal{M}_{ext}^{\mu\nu}$ and  $q_\nu \mathcal{M}_{int}^{\mu\nu} = - q_\nu \mathcal{M}_{ext}^{\mu\nu}$. 

The most general symmetric tensor structure for $B_{\mu\nu}$ takes the form
\begin{align}
B_{\mu\nu} = \sum_{i,j = 1}^{n} C_{ij}(q) k_{i\mu}^\prime k_{j\nu}^\prime + \sum_{i=1}^{n} D_i(q) (k_{i\mu}^\prime q_\nu + k_{i\nu}^\prime q_\mu) + E(q) \eta_{\mu\nu}\, .
\end{align}
The $k_1^\prime,\dots , k_{n}^\prime$ dependence of $C_{ij}(q)$, $D_i(q)$, and $E(q)$ are left implicit. There are no singularities in $q$ since we are assuming $H_{\mu\nu}$ contains only hard exchanges, which enables us to Taylor expand it in powers of $q$.

Taylor expanding $B_{\mu\nu}$ in powers of $q$, we obtain
\begin{align}
B_{\mu\nu} = B^{(0)}_{\mu\nu} + B^{(1)}_{\mu\nu}\, ,
\end{align}
where
\begin{align}
B^{(0)}_{\mu\nu} &= \sum_{i,j=1}^{n} C_{ij}(0)k_{i\mu}^\prime k_{j\nu}^\prime + E(0)\eta_{\mu\nu}\, , \nn \\
B^{(1)}_{\mu\nu} &= \sum_{i,j=1}^{n} q^\alpha \frac{\partial C_{ij}}{\partial q^\alpha}\bigg|_{q=0} k_{i\mu}^\prime  k_{j\nu}^\prime + \sum_{i=1}^{n} D_i(0)(k_{i\mu}^\prime q_\nu + k_{i\nu}^\prime q_\mu) + q^\alpha \frac{\partial E}{\partial q^\alpha}\bigg|_{q=0}\eta_{\mu\nu}\, .
\end{align}
$B_{\mu\nu}$ must be gauge invariant at each order in $q$. Therefore,
\begin{align}
q^\mu B^{(0)}_{\mu\nu} &= q^\nu B^{(0)}_{\mu\nu} = 0 \nn \\
q^\mu B^{(1)}_{\mu\nu} &= q^\nu B^{(1)}_{\mu\nu} = 0 \, .
\end{align}
Since $B^{(0)}_{\mu\nu}$ has no $q$ dependence, this implies that  $B^{(0)}_{\mu\nu} = 0$. On the other hand, $B^{(1)}_{\mu\nu}$ does depend on $q$, so the implication is not immediate. We can, however, rewrite $B^{(1)}_{\mu\nu}$ as follows
\begin{align}
B^{(1)}_{\mu\nu} = q^\alpha B^{(1)}_{\alpha\mu\nu}\, ,
\end{align}
with
\begin{align}
B^{(1)}_{\alpha\mu\nu} = \sum^{n-1}_{i,j=1} \frac{\partial C_{ij}}{\partial q^\alpha}\bigg|_{q=0} k_{i\mu}^\prime  k_{j\nu}^\prime + \sum_{i=1}^{n-1} D_i(0)(\eta_{\mu\alpha}k_{i\nu}^\prime + \eta_{\nu\alpha} k_{i\mu}^\prime) + \frac{\partial E}{\partial q^\alpha} \bigg|_{q=0} \eta_{\mu\nu} \, .
\end{align}
Notice that unlike $B^{(1)}_{\mu\nu}$, $B^{(1)}_{\alpha\mu\nu}$ is independent of $q$. Now, by gauge invariance, we know that
\begin{align}
q^\alpha q^\mu B^{(1)}_{\alpha\mu\nu} = 0\, .
\end{align}
Considering each value of $\nu$ independently, the above condition implies that $B^{(1)}_{\alpha\mu\nu}$ is antisymmetric in its first two indices, that is
\begin{align}
B^{(1)}_{\alpha\mu\nu} = - B^{(1)}_{\mu\alpha\nu}\,.
\end{align}
Of course, the same argument yields the analogous antisymmetry property in the first and third indices. Further, we recall that $B_{\mu\nu}$ is symmetric in $\mu$ and $\nu$, a property which necessarily makes $B^{(1)}_{\alpha\mu\nu}$ symmetric in its second and third indices. Combining all of these properties, we obtain the following chain of interchanges of indices
\begin{align}
B^{(1)}_{\alpha\mu\nu} = -B^{(1)}_{\mu\alpha\nu} = -B^{(1)}_{\mu\nu\alpha} = B^{(1)}_{\nu\mu\alpha} = -B^{(1)}_{\alpha\mu\nu}\, ,
\end{align}
from which we deduce that $B^{(1)}_{\alpha\mu\nu}=0$ and hence $B_{\mu\nu}$ vanishes up to order $O(q^1)$. Our results Eqs.\ \eqref{eq:radiative_hard_scalars} and \eqref{eq:radiative_hard_part_fermions} are therefore the full graviton emission amplitudes from the hard parts with external scalars and fermions respectively, even in the off-shell scenario. Note that the entire graviton emission amplitude is gauge invariant, but the presence of singular terms in $q$ prevents the argument we have just described from showing it has to vanish as well.

\subsection{Graviton Emission from Non-Leading Jets} \label{subsec:graviton_emission_non_leading}

Having discussed the derivation of the radiative hard part for diagrams with leading jets only, we move on to graviton emission from non-leading jets. Up to $O(\lambda^2)$, the non-leading factorized contributions to the elastic amplitude, as well as the leading term, are all gathered in Eq.\ \eqref{eq:elastic_factorized}. The factorized contributions to the elastic amplitude at $O(\lambda^3)$ and $O(\lambda^4)$ are listed in Tables \ref{tab:gamma_3_factorized}, \ref{tab:gamma_4_factorized}, and \ref{tab:exceptional_contributions}.

Consider the contributions appearing in Eq.\ \eqref{eq:elastic_factorized}. To generate radiative contributions from these factorized terms, we need to consider attaching a graviton to each factor separately: the jet functions, the soft cloud, and the hard part. This results in the generic radiative amplitude,
\begin{align} \label{eq:generic_radiative}
\mathcal{M}_{\mu\nu} &= \sum_{i=1}^n \left(\prod_{j\neq i} J_j^f \right) J_{i,\mu\nu}^f  \otimes H + \left(\prod_{i=1}^n J_i^f \right) \otimes H_{\mu\nu} \nonumber \\ 
& \quad +\sum_{\theta \in \Theta_1} \sum_{i=1}^n \left[   \left(\prod_{j\neq i} J_j^f \right) J_{i,\mu\nu}^\theta \otimes H_i^\theta + \sum_{l \neq i } \left( \prod_{j \neq i, l} J_j^f\right) J_{l,\mu\nu}^f J_i^\theta \otimes H_i^\theta +\left(\prod_{j\neq i} J_j^f \right) J_{i}^\theta \otimes H_{i,\mu\nu}^\theta  \right] \nonumber \\
&\quad + \sum_{\theta \in \Theta_2} \sum_{1 \leq i < j \leq n} \bigg[ \left(\prod_{l \neq i,j} J_l^f \right) J_{i,\mu\nu}^\theta J_j^\theta \,  S^\theta \, \otimes  H_{ij}^{\theta\theta}  + \left(\prod_{l \neq i,j} J_l^f \right) J_i^\theta J_{j,\mu\nu}^\theta \,  S^\theta \, \otimes  H_{ij}^{\theta\theta}\nonumber \\
&\quad\quad \quad \quad \quad \quad  + \sum_{h \neq i,j} \left(\prod_{l \neq i,j,h} J_h^f\right) J_i^\theta J_j^\theta J_{h,\mu\nu}^f \,  S^\theta \, \otimes  H_{ij}^{\theta\theta} +  \left(\prod_{l \neq i,j} J_l^f \right) J_{i}^\theta J_j^\theta \,  S_{\mu\nu}^\theta \,  \otimes H_{ij}^{\theta\theta}   \nonumber \\
&\quad\quad \quad \quad \quad \quad  +  \left(\prod_{l \neq i,j} J_l^f \right) J_{i}^\theta J_j^\theta \,  S^\theta \,  \otimes H_{ij,\mu\nu}^{\theta\theta}   \bigg] \nonumber \\
&\quad + O(\lambda)\, ,
\end{align}
where $\Theta_1 = \{fs, fss, fff, f\partial s\}$ and $\Theta_2 = \{fs, f\softs\}$ are two sets of jet labels. We define $S^\theta \equiv 1$ if $\theta = fs$, and $S^\theta \equiv S_{ij}$ if $\theta = f\softs$. Also, we set $S^{\theta}_{\mu\nu} \equiv 0$ if $\theta = fs$, and $S^\theta_{\mu\nu} = S_{ij,\mu\nu}$ if $\theta = f\softs$.  As in Sec.\ \ref{subsec:elastic_factorization},  in the jet label, the superscripts $f$, $s$, and $\softs$ stand for a collinear fermion/antifermion, a collinear scalar, and a soft scalar respectively. The $\partial$ symbol in a jet label refers to the higher dimensional jet function obtained when expanding the hard part in the transverse component of a loop momentum.

\begin{figure}[h!]
\centering
\includegraphics[width = 0.3\textwidth]{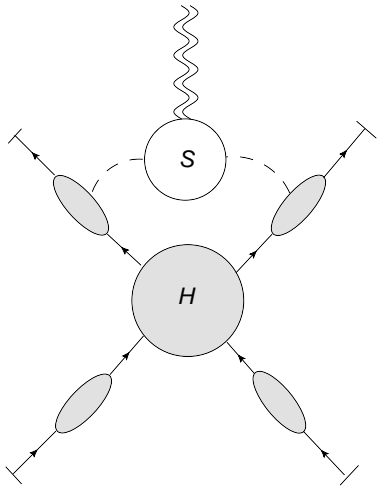}
\caption{Since the graviton can couple to scalars, it is possible to emit a graviton from the soft cloud at $O(\lambda^2)$.}
\label{fig:graviton_soft_cloud_emission}
\end{figure}

To derive the radiative hard parts, we need to apply the general off-shell Ward identity expressed diagrammatically in Fig.\ \ref{fig:general_off_shell_ward} of Appendix \ref{sec:off_shell_ward}. This was done in Ref.\ \cite{Gervais:2017yxv} in the case of photons. The main difference in this case is that gravitons can also couple to scalars, and in particular, it is possible to emit a graviton from the soft cloud, as shown in Fig.\ \ref{fig:graviton_soft_cloud_emission}. This diagram is, in fact, the only instance of graviton emission from a soft cloud to $O(q)$. All other diagrams with a soft cloud identified in Sec.\ \ref{subsec:power_counting} are of order $O(\lambda^3)$ or $O(\lambda^4)$. Since attaching a soft graviton to the soft cloud does not lower the scaling power $\gamma$ of the diagram (see Sec.\ \ref{subsec:graviton_emission_power_counting}), emitting a graviton from these diagrams would not generate a diagram of order $O(\lambda^2)$ or less, as is required to contribute to the soft graviton theorem. The possibility of emitting a graviton from a soft cloud makes contact with the emission of soft gluons in gauge theory, which is the subject of ongoing work, some of which is presented in \cite{Bonocore:2015esa, Laenen:2010uz, Luna:2016idw}.


Applying the off-shell Ward identity to \eqref{eq:generic_radiative} in order to deduce the radiative hard parts yields formulas of the form
\begin{align}\label{eq:non_leading_hard}
H^{\theta}_{\mu\nu} = \mathcal{O}_{\partial, \sigma^{\mu\nu}} H^{\theta} \, ,
\end{align}
where $\mathcal{O}_{\partial, \sigma^{\mu\nu}} $ is an operator built from $\sigma^{\mu\nu}$ and derivatives with respect to external momenta. This is similar to Eqs.\ \eqref{eq:radiative_hard_scalars} and \eqref{eq:radiative_hard_part_fermions}, although the number of terms will be greater due to the increased number of collinear and soft legs in non-leading jets.

For factorized contributions of order higher than $O(\lambda^2)$, as those listed in Tables \ref{tab:gamma_3_factorized} to \ref{tab:exceptional_contributions}, one need only consider graviton emission from the jets, since emitting a graviton from the soft cloud or hard part would leave the scaling power of the diagram unchanged, and hence yield a contribution of order too high to correct the soft graviton theorem.

\subsection{Low Energy Limit} \label{subsec:graviton_low_energy}

Suppose that we are interested in graviton emission in the regime where Low's theorem applies. Specifically, we consider the case of external fermions of mass $m\neq 0$ and take the limit $q\ll \frac{m^2}{E}$, where $E$ is the center of mass energy. Then the elastic amplitude may be expanded in $q$ when applying Low's original analysis and there is no need to consider graviton emission from the jets $J^i(k_i)$. We can simply split the amplitude $\mathcal{M}_{\mu\nu}(k_1...k_n,q)$ into the emission from the external and the internal lines. In \eqref{eq:factorized_graviton_emission}, this corresponds to making the replacement $H \mapsto \mathcal{M}_{el} (k_1\dots k_n)$. Because we have massless scalars, however, the vertex function through which gravity couples to matter has a branch cut starting at $q = 0$. This branch cut is associated with the annihilation of the graviton into two soft scalars and will be shown explicitly in Sec.\ \ref{subsec:off_shell_emission}. The result of these prescriptions, then,  is the expansion
\begin{align}
\label{eq:verylowqlimit}
\mathcal{M}_{\mu\nu}(k_1,\dots, k_n,q) = &\sum_{i=1}^n\frac{\mathcal{V}^{ffG}_{\mu\nu}(k_i, q)}{2 k_i\cdot q + q^2}\left( 1+ q^\rho\frac{\partial}{\partial k_i^\rho} + \frac{1}{2} q^\rho q^\sigma\frac{\partial^2}{\partial k_i^\rho \partial k_i^\sigma}\right) \mathcal{M}_{el} \nonumber \\
&+ \eta_{\mu\nu} \mathcal{M}_{el}  -\frac{1}{2} \sum_{i=1}^n \left(k_{i\mu}\frac{\partial}{\partial k_{i}^\nu} + k_{i\nu}\frac{\partial}{\partial k_i^\mu}\right)\mathcal{M}_{el} \nonumber\\
& +\frac{1}{2}\sum_{i=1}^n \left( (k_i\cdot q) \frac{\partial^2}{\partial k_i^\mu \partial k_i^\nu} - k_{i\mu} q^\rho \frac{\partial^2}{\partial k_i^\rho \partial k_i^\nu}- k_{i\nu} q^\rho \frac{\partial^2}{\partial k_i^\rho \partial k_i^\mu} \right.
\nonumber\\
& \quad \quad \quad \quad \quad \left.   + q^\rho \sigma^i_{\rho \mu} \frac{\partial }{\partial k_{i}^{\nu}} + q^\rho \sigma^i_{\rho \nu} \frac{\partial }{\partial k_{i}^\mu} \right)\mathcal{M}_{el}\, ,
\end{align}
in which all $q$-dependence is explicit to $O(q)$ and the $\xi_j$'s have been omitted for brevity. In the external emission, the function $\mathcal{V}^{ffG}_{\mu\nu}$ combines the external spinor, the fully dressed fermion-graviton vertex, and the numerator of the external fermion propagator. At low energies, we have that $\lambda = m/E \sim 1$ and thus a new small scale lambda must be identified if we are to retain the soft scaling $q = O(\lambda^2)$. In this case, $q << m/E$ and the conditions for applying Low's original analysis are met.

Recalling the definition of the angular momentum operator, (\ref{eq:orbital}), we can rewrite Eq.\ \eqref{eq:verylowqlimit} as the basic CS result, Eqs.\ \eqref{eq:CSresult}-\eqref{eq:orbital}, provided we replace the external emission function $\mathcal{V}^{ffG}_{\mu\nu}$ by its tree level expression and consider an on-shell physical graviton. In particular, the CS formula does receive loop corrections even in the low energy regime.  Note that for fermions, to obtain the ``double $\mathcal{J}$'' form at $O(q)$ explicitly, we require a term of the form $q^\rho q^\sigma \sigma^{(i)}_{\rho\mu} \sigma^{(i)}_{\sigma\nu} \mathcal{M}_{el}$ which is both gauge invariant and non-singular.  We have shown that such terms cannot be part of $H_{\mu\nu}$.  A simple exercise, however, shows that
\begin{align}
q^\rho q^\sigma ( \sigma^{(i)}_{\rho\mu} \sigma^{(i)}_{\sigma\nu} +  \sigma^{(i)}_{\rho\nu} \sigma^{(i)}_{\sigma\mu}) = \frac{1}{2}(q_\mu q_\nu - q^2 \eta_{\mu\nu}) \, ,
\end{align}
which we will see occurs as a gauge invariant contribution to the external amplitude. This allows us to make contact with the CS result shown in Eqs. \eqref{eq:CSresult}-\eqref{eq:orbital}.


\section{External Emission} \label{sec:external_emission}

\subsection{KG Decomposition} \label{subsec:kg_decomposition}

Having investigated the low energy region and the relation of our results \eqref{eq:radiative_hard_scalars} and \eqref{eq:radiative_hard_part_fermions} to the CS formula \eqref{eq:CSresult}-\eqref{eq:orbital}, we now return to the high energy region. In this section, we will be specifically interested in the structure of graviton emission from the jet functions.

Following Del Duca \cite{DelDuca:1990gz} and drawing inspiration from Grammer and Yennie's decomposition \cite{Grammer:1973db}, it is possible to separate the radiative jet functions into a gauge invariant piece and a leading part that obeys the off-shell Ward identity \eqref{eq:JetWard}. For the purpose of analysis, we introduce the notation
\begin{align}
J^{f,\mu\nu}(p,q) &= J^{f,\mu\nu}_L(p,q) + J^{f,\mu\nu}_T(p,q)\, ,
\end{align}
where for simplicity we have made the replacement $k_i \mapsto p$ and opted to use the leading fermionic jet as an illustrative example. The extension of our analysis to other jets follows the same line of reasoning. The jet functions $J^{f,\mu\nu}_L$ and $J^{f,\mu\nu}_T$ satisfy,
\begin{align} \label{eq:J_L_T_def}
q_\mu J^{f,\mu\nu}_L(p,q) &= J^f (p) \left( p^\nu - \frac{1}{2} q_\mu \sigma^{\mu\nu} \right) \nonumber \\
q_\mu J^{f,\mu\nu}_T(p,q) &= 0 \, .
\end{align}
Now, in the spirit of Del Duca's analysis \cite{DelDuca:1990gz}, we introduce projection operators $\tensor{K}{^\rho^ \sigma_\mu_\nu}$ and $\tensor{G}{^\rho^\sigma_\mu_\nu}$ that isolate $J^{\mu\nu}_L$ and $J^{\mu\nu}_T$. Explicitly, this translates into the requirements
\begin{align}\label{eq:KG_defining_property}
J^f_{\rho\sigma}(p,q) \tensor{K}{^\rho^\sigma_\mu_\nu} &= J^f_{L,\mu\nu}(p,q) \, ,\nonumber \\
J^f_{\rho\sigma}(p,q) \tensor{G}{^\rho^\sigma_\mu_\nu}&= J^f_{T,\mu\nu} (p,q) \, .
\end{align}
Of course, this definition is ambiguous as one could always add a transverse piece to $J^{f,\mu\nu}_L$ while maintaining these conditions. 

To identify $K$ and $G$, we first observe that they must add to a product of Kronecker deltas,
\begin{align}
 \delta^\rho_\mu \delta^\sigma_\nu &= \KK + \GG \, .
\end{align}
Drawing inspiration from Del Duca's tensor \cite{DelDuca:1990gz},
\begin{align}
\tensor{G}{^\rho_\mu} &= \delta^\rho_\mu - \frac{q^\rho p_\mu}{p\cdot q} \, ,
\end{align}
we are led to decompose $\delta^\rho_\mu \delta^\sigma_\nu$ as follows,
\begin{align}
\delta^\rho_\mu \delta^\sigma_\nu &= \left(\frac{q^\rho p_{\mu}}{p\cdot q}+\left(\delta^\rho_\mu -\frac{q^\rho p_{\mu}}{p\cdot q} \right)\right)\times \left(\frac{q^\sigma p_{\nu}}{p\cdot q}+\left(\delta^\sigma_\nu -\frac{q^\sigma p_{\nu}}{p \cdot q} \right)\right)\nonumber \\
&= \left(\frac{q^\rho p_{\mu}}{p\cdot q}+\tensor{G}{^\rho_\mu}\right)\left(\frac{q^\sigma p_{\nu}}{p\cdot q}+\tensor{G}{^\sigma_\nu}\right)\, .
\end{align}
Expanding the above product invites us to define
\begin{align}
\GG &= \tensor{G}{^\rho_\mu} \tensor{G}{^\sigma_\nu}\,, \nonumber \\
\KK &= \frac{q^\rho p_\mu}{ p\cdot q} \frac{q^\sigma p_\nu}{ p \cdot q} + \frac{q^\sigma p_\nu}{ p \cdot q} \tensor{G}{^\rho_\mu} + \frac{q^\rho p_\mu}{p\cdot q} \tensor{G}{^\sigma_\nu}\,.
\end{align}
Notice that since $q^\mu \tensor{G}{^\rho_\mu} = 0$, we necessarily have that $\GG$ is transverse to $q$ when contracted with the free indices of $J^f_{T,\mu\nu}$ from Eq.\ \eqref{eq:KG_defining_property}. Thus, $\GG$ has the right property for isolating a transverse component of the jet function $J^f_{\mu\nu}$. For completeness, we should also verify that $J^f_{\rho\sigma}(p,q) \KK$ satisfies the off-shell Ward identity \eqref{eq:JetWard}. We begin with an application of the off-shell Ward identity to rewrite the product $J^f_{\rho\sigma}(p,q) \KK$ in the form
\begin{align} \label{eq:K_graviton_expanded}
J^f_{\rho\sigma}(p,q) \KK &= J^f (p) \left( p_\mu p_\nu - \frac{1}{2} p_\mu  q^\alpha \sigma_{\alpha\nu} - \frac{1}{2} p_\nu q^\alpha \sigma_{\alpha \mu} \right) \frac{1}{p\cdot q}\, ,
\end{align}
from which it immediately follows that
\begin{align}
q^\mu J^f_{\rho\sigma}(p,q) \KK = J^f(p) \left( p_\nu - \frac{1}{2} q^\alpha \sigma_{\alpha \nu} \right) \, .
\end{align}

Suppose now that we have another set of tensors $\TKK$ and $\TGG$ that split $J^f_{\mu\nu}(p,q)$ into a piece obeying the off-shell Ward identity and another piece transverse to $q^\mu$. Then we have
\begin{align}
q^{\mu} J^f_{\rho\sigma}(p,q)(\KK - \TKK) = 0 \, ,
\end{align}
and therefore the difference between $\KK$ and $\TKK$ can be absorbed into $\GG$. Hence, the tensors $\KK$ and $\GG$ are unique up to the addition of a transverse piece to $J^{f,\mu\nu}_L$ and $J^{f,\mu\nu}_T$. 

The application of parity to $J^{f,\mu\nu}(p,q)$ shows that the only gamma matrix structures that can be used in its construction are $1$, $\gamma^\mu$, and $\sigma^{\mu\nu}$. The most general expression for the transverse radiative jet function is then
\begin{align}
J^{f,\mu\nu}_T(p,q) = & \bar{u}(p) \bigg[ F_1 (\eta^{\mu\nu} q^2 - q^\mu q^\nu) \nonumber \\
&+ F_2 ( \eta^{\mu\nu} q^2 - q^\mu q^\nu ) \qslash \nonumber \\
& +  F_3 \left(p^\mu q^\nu + q^\mu p^\nu - \frac{q^2}{p\cdot q} p^\mu p^\nu - \frac{p \cdot q}{q^2} q^\mu q^\nu \right) \nonumber \\
& + F_4 \left((p^\mu q^\nu +  p^\nu q^\mu  - 2\frac{p\cdot q}{q^2} q^\mu q^\nu ) \qslash - (q^2 p^\mu - p\cdot q \, q^\mu ) \gamma^\nu - (q^2 p^\nu - p \cdot q \, q^\nu ) \gamma^\mu\right) \nonumber \\
& + F_5 \left( \left( p^\mu p^\nu - \left(\frac{p\cdot q}{q^2}\right)^2 q^\mu q^\nu \right) \qslash - \frac{p\cdot q}{q^2} ( q^2 p^\mu - p\cdot q \, q^\mu ) \gamma^\nu - \frac{p\cdot q}{q^2} ( q^2 p^\nu - p\cdot q \, q^\nu ) \gamma^\mu \right) \nonumber \\
& + F_6 ( ( q^2 p^\mu - p \cdot q \, q^\mu ) q_\alpha \sigma^{\alpha \nu} +( q^2 p^\nu - p\cdot q \, q^\nu ) q_\alpha \sigma^{\alpha \mu} ) \bigg] \, . \label{eq:Transverse_Pieces}
\end{align}
Each of the coefficients $F_1, \dots, F_6$ multiplies one of the allowed independent symmetric transverse structures that we can build from $\eta^{\mu\nu}$, $p^\mu$, $q^\mu$, $\gamma^\mu$, and $\sigma^{\mu\nu}$.  
Although this is a relatively long list, we are only interested in those terms that vanish no faster than $O(q)$.   The dimensions of the form factors $F_i$ vary from term to term, but, combined with their corresponding tensors, they can have at worst a single algebraic pole, $1/(p\cdot q)$, and no pole in $q^2$.   These restrictions follow immediately from the presence of divergences that are at worst logarithmic for the non radiative jet functions in the limit of zero masses.    Dimensional analysis of the terms in Eq.\ (\ref{eq:Transverse_Pieces}) shows that for $q^2=0$, none of these contributions can appear at order $q$.   For the off-shell case, $q^2\ne 0$, the $F_1$ term may appear, and we will give a one-loop example of how it occurs below.

It is possible to rewrite the transverse contributions in an interesting way if we contract them with an arbitrary graviton polarization tensor $\tilde{h}^{\mu\nu}$. The result is 
\begin{align}\label{eq:Riemanncoupling}
J^f_{\rho\sigma}(p,q) \GG \tilde{h}^{\mu\nu} &= \frac{p_\mu  p_\nu }{(q\cdot p)^2} J^f_{\rho\sigma}(p,q)   (q^\mu q^\nu \tilde{h}^{\rho\sigma} - q^\mu q^\sigma \tilde{h}^{\rho\nu} - q^\nu q^\rho \tilde{h}^{\mu\sigma} + q^\rho q^\sigma \tilde{h}^{\mu\nu}) \, .
\end{align}
We recognize the quantity between brackets on the right hand side as the Riemann curvature tensor (as has also recently been found in \cite{Laddha:2017ygw}) of a plane gravitational wave with polarization vector $\tilde{h}^{\mu\nu}$,
\begin{align} \label{eq:linearized_riemann_def}
R^{\rho\mu\sigma\nu}(\tilde{h}, q) \equiv q^\mu q^\nu \tilde{h}^{\rho\sigma} - q^\mu q^\sigma \tilde{h}^{\rho\nu} - q^\nu q^\rho \tilde{h}^{\mu\sigma} + q^\rho q^\sigma \tilde{h}^{\mu\nu} \, .
\end{align}
The Riemann tensor term of \eqref{eq:Riemanncoupling} is unique among high energy power corrections. It is the only high energy correction that is sensitive to the analytic structure of leading jets. The corrections associated with non-leading jets are given by the $K$ projection and factor from the leading jets as in Eq. \eqref{eq:non_leading_hard}.

\newpage

Combining Eqs.\ \eqref{eq:radiative_hard_part_fermions}, \eqref{eq:K_graviton_expanded}, \eqref{eq:Riemanncoupling}, and \eqref{eq:linearized_riemann_def}  enables us to rewrite the soft graviton theorem for fermions, with only leading fermionic jets taken into account, as
\begin{align}\label{eq:superresult}
\tilde{h}^{\mu\nu}& \mathcal{M}_{\mu\nu}  \nonumber \\
=& \jProd \otimes \sum_{i=1}^n \frac{\tilde{h}^{\mu\nu}}{k_i \cdot q} \left(k_{i\mu} k_{i\nu} - \frac{1}{2} k_{i\mu} q^\rho \sigma^{i}_{\rho\nu} - \frac{1}{2} k_{i\nu} q^\rho \sigma^i_{\rho\mu} \right) H(k_1, \dots, k_i + q, \dots, k_n)  \nonumber \\
&+\jProd \otimes \left[\eta_{\mu\nu}\tilde{h}^{\mu\nu}\, H - \frac{1}{2} \tilde{h}^{\mu\nu} \sum_{i=1}^{n} \left(k_{i\mu}^\prime \frac{\partial H}{\partial k_i^{\prime\nu}} + k_{i\nu}^\prime \frac{\partial H}{\partial k_i^{\prime\mu}}\right)  \right] \nonumber\\
&+\frac{1}{2}\jProd \otimes \sum_{i=1}^{n} \tilde{h}^{\mu\nu} \left(k_i^\prime \cdot q \frac{\partial^2 H }{\partial k_i^{\prime\mu} \partial k_i^{\prime\nu}} - k_{i\mu}^\prime q^\rho \frac{\partial^2 H }{\partial k_i^{\prime\rho} \partial k_i^{\prime\nu}} - k_{i\nu}^\prime q^\rho \frac{\partial^2 H }{\partial k_i^{\prime\rho} \partial k_i^{\prime\mu}} \right. \nonumber \\
& \quad \quad \quad \quad \quad  \quad \quad \quad \quad \quad  \quad \quad \quad  \left. + q^\rho \sigma^{i}_{\rho \mu} \frac{\partial H}{\partial k_i^{\prime\nu}} + q^\rho \sigma^{i}_{\rho \nu} \frac{\partial H}{\partial k_i^{\prime\mu}} \right) \nonumber \\
& +  \sum_{i=1}^n \left(\prod_{j\neq i} J^{f}(k_j)\right) \frac{k_{i\mu} k_{i\nu}}{(k_i \cdot q)^2}  R^{\rho\mu\sigma\nu}(\tilde{h}, q) J^f_{\rho\sigma}(k_i,q) \otimes H \nonumber \\
& + \jProd \otimes \tilde{h}^{\mu\nu} \left[\eta_{\mu\nu} \sum_{i=1}^n \xi_i \cdot \frac{\partial H}{\partial k_i^\prime}  -  \frac{1}{2} \sum_{i,l =1}^n \xi_i \cdot \frac{\partial }{\partial k_i^\prime} \left( k_{l\mu}^{\prime} \frac{\partial H}{\partial k_{l}^{\prime\nu}} + k_{l\nu}^{\prime} \frac{\partial H}{\partial k_{l}^{\prime\mu}} \right) \right]\, .
\end{align}
The first three factorized sums in our result depend only on the non-radiative jet functions and the hard sub-amplitude. This dependence is completely dictated by the off-shell Ward identity, and is consistent with the CS result, Eqs.\ (\ref{eq:CSresult})-\eqref{eq:orbital}  \cite{Cachazo:2014fwa}. The fourth sum organizes contributions that are transverse, and do not follow directly from the Ward identities (\ref{eq:JetWard}). These contributions correspond to the result found by Del Duca for Low's theorem in QED \cite{DelDuca:1990gz}. When the polarization tensor describes the radiation of an external background field by some source, as will be illustrated in the forthcoming example, these contributions couple the scattering process to the Riemann tensor of the background field. Note that the $q$ and $\xi_i$ dependence have been left implicit in the external amplitudes. At low energies, the Riemann tensor correction remains and couples to the fully dressed graviton-fermion external vertex.

\subsection{Example of off-shell Emission} \label{subsec:off_shell_emission}

We begin with a few remarks about the case of a physical on-shell graviton. The polarization tensor of the external graviton takes the form $\tilde{h}^{\rho\sigma} = \epsilon^\rho \epsilon^\sigma$, giving
\begin{align}
R^{\rho\mu\sigma\nu} &= (q^\rho \epsilon^\mu - q^\mu \epsilon^\rho)(q^\sigma \epsilon^\nu - q^\nu \epsilon^\sigma) \, .
\end{align}
In the case of scalar theory, the jet function $J^f_{\rho\sigma}$ can only be built from   $k_{i\rho}$, $q_{\rho}$, and $\eta_{\rho\sigma}$, all of which vanish when contracted with $R^{\rho\mu\sigma\nu}k_{i\mu}k_{i\nu}$ -- this critically depends on the graviton being on-shell. Therefore, we conclude that transverse loop corrections are not present in the case of gravity coupled to scalars when emitting an on-shell graviton.

In the case of Yukawa theories, if the outgoing particle is a scalar, then our previous argument for scalar theories still holds. If the outgoing particle is a fermion, then we can decompose the jet function appearing in the product $k_{i\mu} k_{i\nu} J_{\rho\sigma} R^{\rho\mu\sigma\nu}$ as in equation \eqref{eq:Transverse_Pieces}. Once again, the product $R^{\rho\mu\sigma\nu} k_{i\mu} k_{i\nu}$ annihilates all terms appearing in this decomposition and there is no transverse correction to the soft theorem.

One wonders if we can find instances where loop corrections are not annihilated by the Riemann tensor of linearized gravity. From our discussion we know that, at least for Yukawa and scalar theories, we have to consider an off-shell emitted graviton to find such an occurence. This study is the object of the rest of this section.

The most natural setting for the study of off-shell graviton emission is in the scenario where a scattering amplitude takes place in the viscinity of a strong classical source of background gravitational field. We imagine that this source influences the scattering process by exchanging a single soft graviton, as classical fields are made up of highly occupied soft radiation modes. The emission amplitude of a soft graviton by this classical source is denoted $S^{\rho\sigma}$. We further imagine that this classical source is very heavy and graviton emission from it is thus described by the following amplitude, which is really just the stress tensor of a very heavy object whose recoil we neglect,
\begin{align}\label{eq:massive_object}
S^{\alpha\beta}_{\mathrm{HEAVY}}= M^2 \delta^\alpha_0 \delta^\beta_0\, .
\end{align}

The classical source is coupled to the scattering amplitude through an intermediate propagator
\begin{align}
\tensor{\mathcal{P}}{^\mu^\nu_\alpha_\beta} = \frac{i}{2}\frac{\delta^{\mu}_{\alpha}\delta^{\nu}_{\beta}+ \delta^{\mu}_{\beta}\delta^{\nu}_{\alpha} - \eta^{\mu\nu}\eta_{\alpha\beta}}{q^2}\, .
\end{align}
The gravitational field  polarization tensor $\tilde{h}^{\mu\nu}$ from which we built the Riemann tensor then takes the form
\begin{align}
\tilde{h}^{\mu\nu} \equiv \tensor{\mathcal{P}}{^\mu^\nu_\alpha_\beta} S^{\alpha\beta}_{\mathrm{HEAVY}}\, .
\end{align}
The result of our $KG$ decomposition instructs us to use the above polarization tensor as input for the Riemann tensor which appears in our formula for the transverse loop corrections to external graviton emission \eqref{eq:Riemanncoupling}. For concreteness, we focus on the amplitude where a scalar jet exchanges a single graviton with the heavy classical source,
\begin{align}
\mathcal{M}_{ex} &\equiv J^s_{\rho\sigma}(p,q) \GG \tilde{h}^{\mu\nu} = \frac{p_\mu p_\nu}{(p\cdot q)^2} J^s_{\rho\sigma}(p,q) R^{\rho\mu\sigma\nu} \, , \label{eq:Transverse_Corrections}
\end{align}
and simply ignore the hard part and other jets since they play no role in this discussion. The right hand side completely separates information about the scattered particle contained in $\frac{p_\mu p_\nu}{(p\cdot q)^2} J^s_{\rho\sigma}(p,q)$ from the external gravitational field. It will be convenient to introduce a scalar mass $m_s$, which may be thought of as being $O(\lambda^2)$ in the high energy limit, but is otherwise of arbitrary size. Note that in the high energy region, this choice does not alter our power counting rules listed in Table \ref{tab:power_counting}.

\begin{figure}[h!]
\centering
\includegraphics[width = 0.3\textwidth]{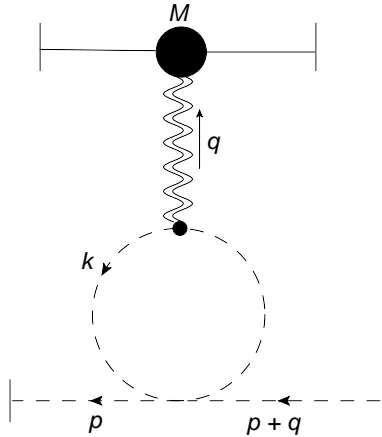}
\caption{The soft graviton is absorbed by a very massive object whose recoil we neglect. This allows us to identify a lowest order correction to the soft graviton theorem in the case of an off-shell soft graviton.}
\label{fig:coupled_tadpole}
\end{figure}

To illustrate how transverse corrections to external emission can be non-vanishing for off-shell gravitons, we will calculate the contribution to $\mathcal{M}_{ex}$ from the diagram shown in Fig.\ \ref{fig:coupled_tadpole}. The amplitude for graviton emission from the jet function in this specific example is found to be
\begin{align}
J^{s,\rho\sigma}(p,q) = (\eta^{\rho\sigma} q^2 - q^\rho q^\sigma)\Theta(p,q)\, ,
\end{align}
with
\begin{align}
\Theta(p,q) = \frac{i g^\prime \mu^{-2\epsilon}\kappa}{(4\pi)^{2-\epsilon}}  \Gamma(\epsilon) \int_0^1 dx \, x(1-x)\left( \frac{m_s^2}{\mu^2}-x(1-x)\frac{ q^2}{\mu^2}\right)^{-\epsilon}\frac{i}{2p\cdot q + q^2} \label{eq:Tadpole_Result} \, ,
\end{align}
and where we have opted to carry out the calculation in $D = 4 - 2 \epsilon$ spacetime dimensions. The constant $g^\prime$ is the coupling of $\phi^4$ theory.


We can then couple this loop correction to the massive object in \eqref{eq:massive_object}, as illustrated in Fig.\ \ref{fig:coupled_tadpole}. This results in the following expression for the single graviton exchange amplitude,
\begin{align} \label{eq:single_graviton_exchange}
\mathcal{M}_{ex} &= \frac{iM^2}{2q^2}[(3-D)q^2 - 2(q_0)^2]\Theta(p,q)\, ,
\end{align}
which is non-vanishing in general.

Consider first the region where Low's analysis is applicable, $q \ll m_s$. Working with the integral representation of the jet function \eqref{eq:Tadpole_Result} provides a more transparent analysis of this region. The integrand can be expanded in powers of $q^2$ using the binomial theorem since $q^2 \ll  m_s^2$, yielding
\begin{align}
\Theta(p,q) &= \frac{i g^\prime \mu^{-2\epsilon} \kappa}{(4\pi)^{2-\epsilon}}\left(\frac{m_s^2}{\mu^2}\right)^{-\epsilon} \Gamma(\epsilon) \left[ \frac{1}{6} + \frac{\epsilon}{30} \frac{q^2}{m_s^2} + O(q^4) \right] \frac{i}{2p\cdot q + q^2} \, .
\end{align}
In an on-shell renormalization scheme, one removes the full correction at $q^2 = 0$. Implementing this scheme leaves us with the following contribution to the jet function,
\begin{align}
\Theta_R(p,q) &= \frac{i g^\prime  \kappa}{(4\pi)^{2}} \left[\frac{1}{30}\frac{q^2}{m_s^2} + O(q^4) \right] \frac{i}{2 p\cdot q + q^2} \, .
\end{align}
We thus obtain a correction of order $q/m_s^2$.

If we return to the region $q = O(m_s)$, then we  resort to fully evaluating the integral in \eqref{eq:Tadpole_Result}, obtaining the result
\begin{align}
\Theta(p,q) &= \frac{i g^\prime \mu^{-2\epsilon} \kappa}{(4\pi)^{2-\epsilon}} \left[\frac{1}{6\epsilon} - \frac{\gamma}{6} + \frac{2}{3} \frac{m_s^2}{q^2} - \frac{1}{6} \log\left(\frac{m_s^2}{\mu^2}\right) \right. \nonumber \\
& \quad \left.+ \frac{1}{6} \left(1+\frac{2m_s^2}{q^2}\right) \sqrt{1-\frac{4m_s^2}{q^2}} \log\left(\frac{\sqrt{1-4m_s^2/q^2}-1}{\sqrt{1-4m_s^2/q^2}+1}\right) + \frac{10}{36}\right]\frac{i}{2p\cdot q + q^2} \, , \label{eq:Tadpole_Explicit}
\end{align}
where we have expanded in $\epsilon$ before performing the integral while avoiding any expansion in $q$. Proceeding with an on-shell renormalization scheme as we did before, we expand the counterterm in powers of $\epsilon$, thereby obtaining
\begin{align}
c.t. &= \frac{i g^\prime \mu^{-2\epsilon} \kappa}{(4\pi)^{2-\epsilon}} \left(\frac{1}{6\epsilon} - \frac{\gamma}{6}  - \frac{1}{6} \log\left(\frac{m_s^2}{\mu^2}\right) \right) \frac{i}{2 p\cdot q + q^2}\, .
\end{align}
Subtracting this counterterm from \eqref{eq:Tadpole_Explicit} results in the correction,
\begin{align}
\Theta_R(p,q) &=  \frac{i g^\prime \kappa}{(4\pi)^{2}} \left[\frac{2}{3} \frac{m_s^2}{q^2}+ \frac{1}{6} \left(1+\frac{2m_s^2}{q^2}\right) \sqrt{1-\frac{4m_s^2}{q^2}} \log\left(\frac{\sqrt{1-4m_s^2/q^2}-1}{\sqrt{1-4m_s^2/q^2}+1}\right) + \frac{10}{36}\right]\frac{i}{2p\cdot q + q^2}\,.
\end{align}
The leading term of this expression is of order $O(q^{-1})$. It is worth noting that all of the apparent poles in this expression cancel.

It is also interesting to take the limit $q \gg m_s$ so that the graviton momentum is no longer soft. Retaining the leading term only, we obtain
\begin{align}
\Theta_R(p,q) &= \frac{i g^\prime \kappa}{(4\pi)^{2}}\frac{1}{6}\log \left(\frac{m_s^2}{-q^2}\right) \frac{i}{2p\cdot q + q^2}\, .
\end{align}
In this case, the soft graviton theorem receives a logarithmic correction from the external jet functions. In all cases, we find that the Riemann tensor is coupled to non-vanishing loop corrections from the jet, whether we are in the high or low energy regime. We conclude by mentioning that the Riemann tensor corrections we have identified can be viewed as quantum corrections to the Newtonian potential \cite{Holstein:2004dn, Bjerrum-Bohr:2016hpa}.


\section{Conclusion}

Following Del Duca \cite{DelDuca:1990gz} and Ref.\ \cite{Gervais:2017yxv}, we have applied power counting techniques to derive an extension of the soft graviton theorem to Yukawa and scalar theories at all loop orders in both the high energy and low energy regions. Our strategy at high energies is to apply Low's original analysis \cite{Low:1958sn} to a radiative amplitude factorized into jets, a soft cloud, and a hard part. This factorization is a solution to the obstacle arising when, in the high energy region, invariants built with the soft graviton momentum become of the same order of magnitude as the other invariants in denominators of the loop integral. This phenomenon occurs in the vicinity of pinch surfaces and prevents us from expanding the elastic amplitude in powers of $q$, as is typically done in an argument \`a la Low. Factorizing the amplitude isolates these non-analytic contributions and encapsulates them into the jet functions and the soft cloud. The hard parts, on the other hand, get their leading contribution from hard exchanges and can legitimately be expanded in powers of $q$.

In the high energy region, the total center of mass energy $E$ is very large compared to the mass $m$ of the fermions. This led us to identify the parameter $\lambda \equiv m/E$ as a small quantity suitable for expressing the orders of magnitude of the various quantities in the problem. In particular, the soft graviton theorem was recast as an expansion in $\lambda$ rather than $q$. We have designed, in Ref.\ \cite{Gervais:2017yxv} and the present paper, a power counting technique that allows us to determine the order of magnitude of any factorized contribution corresponding to a given reduced diagram. We showed that reduced diagrams contributing to the soft graviton theorem are in one-to-one correspondence with the reduced diagrams of the elastic amplitude of orders ranging from $O(\lambda^0)$ to $O(\lambda^4)$. This allowed us to determine all factorized contributions to the radiative amplitude from an analysis of the pinch surfaces of the elastic amplitude. At orders $\lambda^3$ and $\lambda^4$, the reduced diagrams resulting from this analysis are shown in Figs.\ \ref{fig:g_4_partition_4}  to \ref{fig:hard_connection_example}, with the associated factorized contributions listed in Tables \ref{tab:gamma_3_factorized}, \ref{tab:gamma_4_factorized}, and \ref{tab:exceptional_contributions}. The jet functions identified using power counting are reminiscent of the higher dimension operators of the SCET approach to soft theorems \cite{Larkoski:2014bxa}. The jets are also analogous to final state wave functions in bound state scattering \cite{Lepage:1980fj, Lepage:1979zb, Efremov:1979qk}. The hard parts, on the other hand, play a role similar to the matching coefficients of effective field theory. A systematic algorithm for computing them would involve a series of nested subtractions similar to the ``garden and tulip'' construction of \cite{Collins:1981uk}. This is also closely related to the nested subtractions of \cite{Erdogan:2014gha}.

Inspired by Del Duca \cite{DelDuca:1990gz}, we have also applied a decomposition similar to Grammer and Yennie's $GK$ decomposition \cite{Grammer:1973db} to the external amplitude. We found that jet functions supply loop corrections to the soft graviton theorem that are coupled to a Riemann curvature tensor for linearized gravity at low and high energies. In low energy scattering, the Riemann tensor contribution is the only correction to the CS tree level theorem.

Although we have not touched upon the problems of double graviton emission and virtual graviton correction, these can be addressed using our methods. An all loop order analysis of the soft graviton theorem in the presence of gravity loops using our technique is the subject of ongoing work, as are the subjects of photon and gluon emission at all loop orders in gauge theories. For gauge theories, work in this direction has already been presented in \cite{Bonocore:2015esa, Laenen:2010uz, Luna:2016idw, Bonocore:2016awd}.


\section*{Acknowledgements}
The author is grateful to George Sterman for suggesting this problem and for many helpful discussions and suggestions. The author also thanks the Fonds de recherche du Qu\'ebec --  nature et technologies for their financial support through their doctoral research scholarships. This work was supported in part by the National Science Foundation, grants
PHY-1316617 and 1620628.


\appendix

\section{Appendix: Diagrammatic Derivation of the Off-shell Ward identity in Scalar and Yukawa theory} \label{sec:off_shell_ward}

\begin{figure}[h!]
\centering
\includegraphics[width = 1.0 \textwidth]{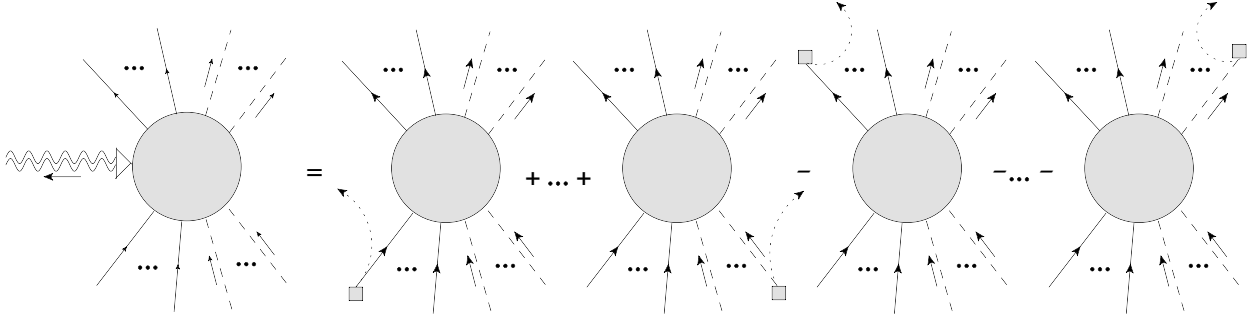}
\caption{The general off-shell Ward identity. This identity holds regardless of whether the external particles are collinear, hard, or soft.}
\label{fig:general_off_shell_ward}
\end{figure}

\begin{figure}[h!]
\centering
\includegraphics[width = 0.7 \textwidth]{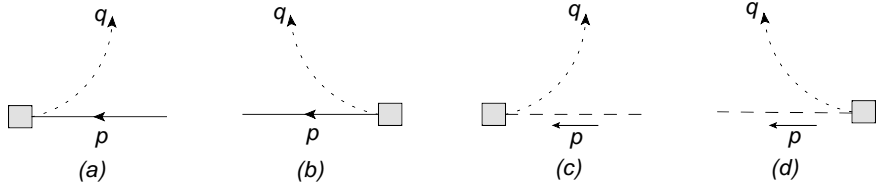}
\caption{The box vertices represent the emission of a ghost graviton after the graviton momentum $q$ is contracted with the radiative amplitude. Vertices (a) and (b) are denoted $W_{fG}^\mu$ and have the same expression. In (a), $W_{fG}$ acts to the left of the fermion propagators, whereas in (b), $W_{fG}$ acts on the right. Likewise, vertices (c) and (d) also have the same value, and are both denoted $W_{sG}^\mu$.}
\label{fig:box_vertices}
\end{figure}

Diagrammatically, the off-shell Ward identity can be expressed as in Fig.\ \ref{fig:general_off_shell_ward} \cite{Brout:1966oea,Bessler:1969py,Coriano:2011zk,Coriano:2012cr}. The box vertices represent the emission of a ``ghost'' graviton -- see Fig.\ \ref{fig:box_vertices}. Their corresponding expressions are given by
\begin{align}
W_{fG}^\nu &= \frac{\kappa}{2} \left(p^\nu - \frac{1}{2} q_\mu \sigma^{\mu\nu}\right) \, , \nonumber \\
W_{sG}^\nu &= \frac{\kappa}{2} p^\nu \, ,
\end{align}
where $\sigma_{\mu\nu} = \frac{1}{4}[\gamma_\mu, \gamma_\nu]$. The vertex $W_{fG}$ acts on the left of a string of fermion propagators if the fermion arrows point towards the box in the corresponding diagram. Conversely, $W_{fG}$ will act to the right of a string of fermionic propagators if the fermion arrow points away from the box in the diagram.


We give a sketch of how the off-shell Ward identity is derived diagrammatically. This proof clearly shows why the Ward identity holds at fixed loop momenta. The key identity we use is the simplest example of the Feynman identity,
\begin{align}\label{eq:simplest_feynman_identity}
q_\mu V_{ffG}^{\mu\nu} = \left(\frac{i\kappa}{4}\right) \bigg( & \frac{1}{2} \gamma^\nu (p^2 - (p+q)^2) \nonumber \\
&+ \frac{1}{2} p^\nu ((1-A) (\pslash - m) - (1+A) ( \pslash + \qslash - m)) \nonumber \\
&+ \frac{1}{2} (p+q)^\nu ( (1+A) (\pslash -m) - (1-A) (\pslash + \qslash - m))\bigg) \, ,
\end{align}
for a fermion emitting a graviton, and
\begin{align}
q_\mu V_{ssG}^{\mu\nu} = \left(\frac{i\kappa}{2}\right) \left((p+q)^\nu p^2 - p^\nu (p+q)^2 \right) \, ,
\end{align}
for a scalar emitting a graviton. The incoming fermion/scalar momentum is taken to be $p+q$ while the outgoing momentum is $p$ --  in Fig.\ \ref{fig:matter_graviton_vertices}, this corresponds to making the replacements $p\mapsto p+ q$ and $p^\prime \mapsto p$. The parameter $A$ can take the value $1$ if we normalize our Lagrangian with the square root of the determinant of the vierbein field $\sqrt{e}$, or $2$ if we opt for $\sqrt{g}$, where $g$ is the determinant of the metric. The change in the normalization of the Lagrangian is compensated by a change in the normalization of the fermion field \cite{BjerrumBohr:2004mz, Veltman:1975vx}.

\begin{figure}[h!]
\centering
\includegraphics[width = 1.0 \textwidth]{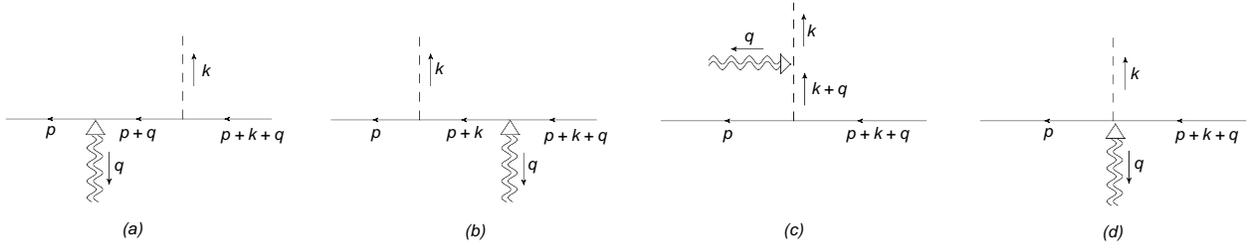}
\caption{The amplitudes in (a), (b), (c), and (d) will be denoted by $q_\mu M_a^{\mu\nu}$, $q_\mu M_b^{\mu\nu}$, $q_\mu M_c^{\mu\nu}$ and $q_\mu M_d^{\mu\nu}$ respectively.}
\label{fig:off_shell_ward_proof}
\end{figure}

To illustrate how these identities are used to prove the off-shell Ward identity, we evaluate the four amplitudes shown in Fig.\ \ref{fig:off_shell_ward_proof}. Note that the ends of the external particle lines are not reduced.


The amplitudes appearing in Fig.\ \ref{fig:off_shell_ward_proof} have the expressions,
\begin{align}
q_\mu M_a^{\mu\nu} = & \frac{i}{\pslash - m} \nonumber \\ 
&\times \left(\frac{i\kappa}{4}\right) \bigg(  \frac{1}{2} \gamma^\nu (p^2 - (p+q)^2 ) \nonumber \\
& + \frac{1}{2} p^\nu ( (1- A) ( \pslash  - m) - (1 + A) ( \pslash + \qslash - m) ) \nonumber \\
& + \frac{1}{2} ( p + q)^\nu ( (1 + A) ( \pslash  -m ) - (1 -A) ( \pslash + \qslash -m )) \bigg) \nonumber \\
&\times  \frac{i}{ \pslash + \qslash - m}   (-i g) \frac{i}{k^2} \frac{i}{ \pslash + \kslash + \qslash - m} \, ,
\end{align}
\begin{align}
q_\mu M_b^{\mu\nu} = & \frac{i}{\pslash - m} (-i g) \frac{i}{k^2} \frac{i}{ \pslash + \kslash - m} \nonumber \\ 
&\times \left(\frac{i\kappa}{4} \right)  \bigg(  \frac{1}{2} \gamma^\nu ((p+k)^2 - (p+k+q)^2 ) \nonumber \\
& + \frac{1}{2} (p + k)^\nu ( (1- A) ( \pslash + \kslash - m) - (1 + A) ( \pslash + \kslash + \qslash - m) ) \nonumber \\
& + \frac{1}{2} ( p + k + q)^\nu ( (1 + A) ( \pslash + \kslash -m ) - (1 -A) ( \pslash + \kslash + \qslash -m )) \bigg) \nonumber \\
&\times \frac{i}{ \pslash + \kslash + \qslash - m} \, ,
\end{align}
\begin{align}
q_\mu M_c^{\mu\nu} =& \frac{i}{\pslash - m} (-i g) \frac{i}{k^2} \left(\frac{i\kappa}{2} \right) \left((k+q)^\nu k^2 - k^\nu (k+q)^2 \right) \frac{i}{(k+q)^2} \frac{i}{\pslash + \kslash + \qslash - m} \, ,
\end{align}
and
\begin{align}
q_\mu M_d^{\mu\nu} &= \frac{i}{\pslash - m} \left(\frac{-i g\kappa}{4}\right) A q^\nu \frac{i}{\pslash + \kslash + \qslash - m} \frac{i}{k^2} \, .
\end{align}

Summing $q_\mu M_a^{\mu\nu}$, $q_\mu M_b^{\mu\nu}$, $q_\mu M_c^{\mu\nu}$, and $q_\mu M_d^{\mu\nu}$, and performing some routine algebra, we obtain 
\begin{align}
\sum_{\theta = a,b,c,d} q_\mu M_\theta^{\mu\nu} &= \nonumber \\
\left(\frac{g \kappa}{4} \right) \bigg( &  \frac{1}{2} ((\pslash + m) \gamma^\nu + \gamma^\nu ( \pslash + \qslash - m) ) \frac{1}{k^2} \frac{1}{\pslash + \qslash - m} \frac{1}{\pslash + \kslash + \qslash - m} \nonumber \\
& +  ( p + \frac{1}{2} (1+A) q)^\nu   \frac{1}{k^2} \frac{1}{\pslash + \qslash - m} \frac{1}{\pslash + \kslash + \qslash - m} \bigg) \nonumber \\
- \left(\frac{g \kappa}{4} \right) \bigg( &  \frac{1}{2} \frac{1}{\pslash  - m}\frac{1}{\pslash + \kslash - m}\frac{1}{k^2}  (\gamma^\nu (\pslash + \kslash + \qslash+ m) + ( \pslash + \kslash - m)\gamma^\nu  )\nonumber \\
& + ( p + k + \frac{1}{2} (1-A) q)^\nu ) \frac{1}{\pslash  - m}\frac{1}{\pslash + \kslash - m}\frac{1}{k^2} \bigg) \nonumber \\
+ \left(\frac{g \kappa}{2} \right) & \frac{1}{\pslash  - m}\frac{1}{\pslash + \kslash + \qslash - m}\frac{(k+q)^\nu}{(k+q)^2} \, .
\end{align}
Using the relation
\begin{align}
(\pslash + m) \gamma^\nu + \gamma^\nu ( \pslash + \qslash - m) &= ( 2 p + q)^\nu + \frac{1}{2} [ \gamma^\nu, \qslash] \, ,
\end{align}
and defining $\sigma^{\mu\nu} \equiv \frac{1}{4} [ \gamma^\mu, \gamma^\nu ]$, we find
\begin{align}
\sum_{\theta = a,b,c,d} q_\mu M_\theta^{\mu\nu} =& \left(\frac{\kappa}{2}\right) \frac{i}{\pslash - m} ( -i g) \frac{i}{\pslash + \kslash - m} \frac{i}{k^2} \left( (p+k)^\nu - \frac{1}{2} q_\mu \sigma^{\mu\nu} \right) \nonumber \\
& -  \left(\frac{\kappa}{2}\right)  \left( (p+ q)^\nu - \frac{1}{2} q_\mu \sigma^{\mu\nu} \right)  \frac{i}{\pslash + \qslash - m} ( -i g) \frac{i}{\pslash + \kslash + \qslash  - m} \frac{i}{k^2} \nonumber \\
& -  \left(\frac{\kappa}{2}\right)  \frac{i}{\pslash - m} ( -i g) \frac{i}{\pslash + \kslash + \qslash  - m} \frac{i}{(k+q)^2} (k + q)^\nu \, ,
\end{align}
where we have set $A=2$. This is exactly the result we expect from the off-shell Ward identity.

To extend our approach to all diagrams, including those with arbitrarily many loops, we need to apply \eqref{eq:simplest_feynman_identity} successively to adjacent graviton insertions along a fermion line. The resulting cancellations along each fermion line will give us  two terms, one with a $+$ sign and an operator $W_{fG}$ acting to the right of the corresponding series of fermion propagators, and one with a $-$ sign and a $W_{fG}$ operator acting to the left of the corresponding series of fermion propagators. If we need to capture nested cancellations along a fermion loop, a shift in loop momentum by $q$ is required. The treatment of scalars is entirely analogous, and is in fact simpler. This diagrammatic argument shows that the Ward identity holds separately for each pinch surface, since we are working at fixed loop momenta, except in the occurrence of a fermion or scalar loop. Note that a shift of the loop momentum by $O(q)$ does not mix collinear and soft loop momenta.

\bibliographystyle{ieeetr}
\bibliography{bib_all_references}

\end{document}